\documentclass[11pt]{article}

\usepackage{amssymb, latexsym, amsmath}
\usepackage{amsfonts, graphics, amstext}
\usepackage{graphicx}
\usepackage{subcaption}
\usepackage{color}
\usepackage{lmodern}
\usepackage{slashed}
\usepackage{overpic}

\usepackage{tikz}
\usetikzlibrary{positioning}

\def\be{\begin{equation}}
\def\ee{\end{equation}}
\def\bea{\begin{eqnarray}}
\def\eea{\end{eqnarray}}
\def\beas{\begin{eqnarray*}}
\def\eeas{\end{eqnarray*}}

\begin{document}

\sloppy

\newtheorem{theorem}{Theorem}[section]
\newtheorem{definition}[theorem]{Definition}
\newtheorem{proposition}[theorem]{Proposition}
\newtheorem{example}[theorem]{Example}
\newtheorem{remark}[theorem]{Remark}
\newtheorem{cor}[theorem]{Corollary}
\newtheorem{lemma}[theorem]{Lemma}

\renewcommand{\theequation}{\arabic{section}.\arabic{equation}}

\title{The transition to phenomenological behaviour of static solutions of the Einstein-Dirac system for an increasing number of fermions}

\author{H{\aa}kan Andr\'{e}asson\\
        Mathematical Sciences\\
        Chalmers University of Technology\\
        University of Gothenburg\\
        SE-41296 Gothenburg, Sweden\\
        email: hand@chalmers.se\\
        \\
        Joakim Blomqvist\\
        Mathematical Sciences\\
        Chalmers University of Technology\\
        University of Gothenburg\\
        SE-41296 Gothenburg, Sweden\\
        email: joablo@student.chalmers.se
        }

\maketitle


\begin{abstract}
Static spherically symmetric solutions to the Einstein-Dirac system were constructed numerically for the first time in 1999 by Finster, Smoller and Yau \cite{FSY1} in the case of two fermions. In 2020 this result was generalized by Leith, Hooley, Horne and Dritschel \cite{LHHD} to a system consisting of an even number $\kappa$ of fermions. They constructed solutions for $2\leq\kappa\leq 90$. The purpose of the present investigation is to compare the properties of static solutions of the Einstein-Dirac system with static solutions of the Einstein,-Vlasov system as the number of fermions increases, that is, for $2\leq\kappa \leq 180$. Since the Einstein\,-Vlasov system is a fully classical physical model, whereas the Einstein-Dirac system is semiclassical and thus has a quantum signature, this framework provides an excellent opportunity to study the transition from quantum to classical behaviour. It turns out that even for a comparatively small number of particles, the features of the solutions are remarkably similar. For both systems, we find highly relativistic solutions having a multi-peak structure with strikingly similar characteristics. We also investigate the maximum compactness ratio $\sup 2m/r$ of the solutions. The solutions of both systems share the fundamental properties regarding the maximum compactness ratio and obey the inequality derived in \cite{A2}. 
Furthermore, we investigate the sign of the pressure components of solutions of the Einstein-Dirac system. For small values of $\kappa$, there are regions where the radial pressure is negative. These regions disappear 
as $\kappa$ increases. This supports the interpretation we make as a transition from quantum to classical behaviour as the number of fermions increases.

\end{abstract}

\section{Introduction}\label{sec-intro}
In 1999 Finster, Smoller and Yau \cite{FSY1} constructed, by numerical methods, spherically symmetric static solutions to the Einstein-Dirac system. These gravitationally localized solutions are often referred to as Einstein-Dirac solitons, with all metric and fermion fields regular at the origin and the resulting spacetime being asymptotically flat. This class of solutions was already considered by Lee and Pang \cite{LP} in 1987 but their analysis was based on approximations. The Dirac equation gives a quantum mechanical description of fermions, whereas the Einstein equations belong to classical physics. The Einstein-Dirac system thus constitutes 
a semiclassical model for the gravitational interaction of fermions. 

To obtain spherically symmetric solutions, Finster et al. considered two neutral fermions with opposite spin so that the total angular momentum vanishes; a necessary property of a spherical symmetric solution. In 2020 this idea was generalized by Leith et al. \cite{LHHD} to a system consisting of many particles or, more precisely, to an even number $\kappa$ of neutral fermions. Their study focuses on the case where $\kappa$ is large compared to the case with two neutral fermions, and values of $\kappa$ up to 90 were considered. Leith et al. find solutions in which strong gravitational effects are in evidence, i.e., exceedingly compact solutions characterized by a very high redshift. Their solutions are soliton-like wave functions which they interpret in terms of a form of self-trapping, where the fermions become localized on shells, which are separated by regions where the energy density almost vanishes. Accordingly, the graph of the energy density, as a function of the radius, becomes multi-peaked. This multi-peak structure of high-redshift solutions of the Einstein-Dirac system was thought to be a phenomenon \cite{LHHD} that had previously not been observed for other specific physical systems (though it had been observed in models with a given equation of state \cite{KRS}). However, a very similar type of solutions was obtained by the first author and Gerhard Rein in 2007 in a study of static solutions of the Einstein\,-Vlasov system \cite{AR}. The purpose of the present work is to make a careful comparison of the static solutions of these two systems as the number of particles $\kappa$ of the Einstein-Dirac system increases. We find that even for a small number of particles the properties of the solutions are notably similar despite the fact that the matter models are so different. Since the Einstein\,-Vlasov system is a fully classical physical model, whereas the Einstein-Dirac system is semiclassical and has a quantum signature, this framework provides an excellent opportunity to study the transition from quantum to classical behaviour. 

A useful quantity for studying the transition to classical behaviour, is the pressure. For phenomenological matter models, such as fluids or Vlasov matter, the pressure is non-negative everywhere, whereas solutions of field-theoretical matter models may have regions where the pressure is negative. In this work, we find indeed that there are negative pressure regions for solutions of the Einstein-Dirac system when $\kappa$ is small. When $\kappa$ increases these regions disappear (for finite $\kappa$) and the pressure becomes non-negative everywhere and we interpret this as a transition to classical behaviour. 
We find it remarkable that this transition occurs for a comparatively small number of fermions. 
Another aspect of negative pressure is that in the literature about the Einstein equations, it is common that the pressure is \textit{chosen} to be negative. A common choice is that the pressure $p=-\rho$ for an isotropic model
where $\rho$ is the energy density, cf. e.g. \cite{MM}. For a genuine coupled Einstein-matter system such as the Einstein-Dirac system or the Einstein\,-Vlasov system, choosing an equation of state such as $p=-\rho$ cannot be made freely. The relation between pressure and energy density is implicit and is only known once a solution has been obtained. The amplitude of the modulus of the pressure in the negative regions is very small for solutions of the Einstein-Dirac system. A similar observation is made in the case of $\ell$-boson stars \cite{AS1}. The pressure in these cases is thus far from fulfilling $p=-\rho$. Hence, making an arbitrary choice of the equation of state may lead to highly unrealistic physical systems.

The overall aim in the present work is to compare several characterizing properties of highly relativistic solutions of the Einstein-Dirac system and the Einstein\,-Vlasov system. In addition to the multi-peak structure and the negative pressure-regions as discussed above, we also compare results on the maximum compactness ratio 
\begin{equation}\label{Gamma}
\Gamma:=\sup \frac{2m(r)}{r},
\end{equation}
which determines the maximum gravitational red-shift. We investigate how well the energy condition 
\begin{equation}\label{energycondition-intro}
p_r+2p_{\perp}\leq \rho,
\end{equation}
which always holds for solutions of the Einstein\,-Vlasov system, is satisfied for solutions of the Einstein-Dirac system. Here, $p_r, p_{\perp}$ and $\rho$ are the radial pressure, the tangential pressure and the energy density, respectively.  
The connection between $\Gamma$ and the energy condition $p_r+2p_{\perp}\leq \rho$ was investigated in \cite{A2}, where it was shown that

\begin{equation}\label{Gamma-intro}
\Gamma\leq \frac89,
\end{equation}
for any solution to the Einstein-matter system for which (\ref{energycondition-intro}) holds and for which the radial pressure is non-negative.
Note that (\ref{Gamma-intro}) is not the standard Buchdahl bound, which requires that the energy density is non-increasing and that the pressure is isotropic, two assumptions which are violated in our case. It will be seen that the solutions of the two systems closely share properties with respect to maximum compactness $\Gamma$ and the energy condition (\ref{energycondition-intro}), as $\kappa$ and the central redshift $z$ increase. In particular, we find that (\ref{Gamma-intro}) also holds for solutions of the Einstein-Dirac system. The question of sharpness of (\ref{Gamma-intro}) is left open in this work: for solutions of the Einstein\,-Vlasov system (\ref{Gamma-intro}) is sharp, but whether this is also the case for solutions of the Einstein-Dirac system is an open question. We do find numerical evidence that sharpness may hold, also for the Einstein-Dirac system, but it is necessary to go beyond $\kappa=180$ to conclude, and this is numerically very demanding. We mention here that the first author, together with Fajman and Thaller, previously observed a similar analogy on the compactness ratio $\Gamma$ between solutions of the Einstein-Maxwell system and the massless Einstein\,-Vlasov system, cf. \cite{W} and \cite{AFT}. Furthermore, the compactness results for the extreme $\ell$-boson stars in \cite{AS1} are also closely related when $l$ is large, as we briefly discuss in Section \ref{sec-sharpness}. 

The outline of the paper is as follows. In the next section, the static spherically symmetric Einstein-Dirac system for an even number $\kappa$ of fermions is formulated where we closely follow \cite{LHHD}. In Section \ref{sec-EV} the analogous equations for the Einstein\,-Vlasov system are given. Some general features of the solutions of the Einstein-Dirac system and the dependence on $\kappa$ and $z$ are given in Section \ref{sec-classify}. In Section \ref{sec-comparison} a comparison between solutions of the Einstein-Dirac system and the Einstein\,-Vlasov system is given in the two cases $\kappa=16$ and $\kappa=90$. The macroscopic properties of the solutions are strikingly similar. The question of negative pressure regions for solutions of the Einstein-Dirac system is investigated in Section \ref{sec-pressure} where we find evidence for a transition to classical behaviour as $\kappa$ increases. In Section \ref{sec-Gamma} the energy condition (\ref{energycondition-intro}) is studied. It is found to be valid for compact solutions of the Einstein-Dirac system, which implies that (\ref{Gamma-intro}) also holds for solutions of the Einstein-Dirac system. The issue of sharpness is discussed in Section \ref{sec-sharpness}. 
A brief discussion on other matter models is also included in this section.

\section{The Einstein-Dirac system}\label{sec-ED}
The purpose of this section is to write down the equations of the Einstein-Dirac system and to introduce the parameters that characterize the solutions we construct. A brief outline on how the equations are obtained is given below; we refer to \cite{LHHD} and \cite{FSY1} for more details. 

We consider a situation where the fermions are arranged in a filled shell with vanishing overall angular momentum, a condition which is necessary for the configuration to be spherically symmetric. The total angular momentum of each individual fermion consists of its spin and its orbital angular momentum, which is taken to be $j\in \{\frac12,\frac32,...\}$. Using the Hartree-Fock formalism, the overall wave function can be written as
\[
\Psi=\Psi_{j,k=-j}\wedge \Psi_{j,k=-j+1}\wedge ... \wedge \Psi_{j,k=j}, 
\]
where $\Psi_{j,k}$ is the wave function of an individual fermion with angular momentum component in the $z$-direction equal to $k$. The number of fermions in a filled shell is accordingly $\kappa:=2j+1$. In this work, we consider cases up to $\kappa=180$, which can be compared with the work \cite{LHHD} where $\kappa\leq 90$. It should be clarified that severe numerical difficulties arise as $\kappa$ increases. 
In natural units $\hbar=c=G=1$ the Einstein-Dirac system reads 
\begin{equation}
R_{\mu \nu} - \frac{1}{2}Rg_{\mu \nu} = 8\pi T_{\mu \nu},\nonumber
\end{equation}
with Dirac's equation
\begin{equation}
  (i\slashed{D} - m)\Psi = 0.\nonumber
\end{equation}
Here $m$ is the mass of each fermion and $\slashed{D}$ is the Dirac operator 
\[
\slashed{D}=i\gamma^{\mu}(\partial_{\mu}+\Gamma_{\mu}).
\]
$\Gamma_{\mu}$ is the spin connection and $\gamma^{\mu}$ are the Dirac gamma matrices generalized to curved spacetimes so that $\{\gamma^{\mu},\gamma^{\nu}\}=-2g^{\mu\nu}$. $T_{\mu \nu}$ is obtained from $\Psi$ and the metric, cf. \cite{FSY1}. The non-trivial components of $T_{\mu \nu}$ are explicitly given below in terms of the radial pressure $p_r$, the tangential pressure $p_{\perp}$, and the energy density $\rho$, respectively. 
To write down the Einstein-Dirac system in coordinates we follow \cite{FSY1,LHHD} and write the metric as
\begin{equation}\label{metric-ED}
ds^2=-T^{-2}(r)dt^2+A^{-1}(r)dr^2+r^{2}(d\theta^{2}+\sin^{2}{\theta}d\varphi^{2}),
\end{equation}
where $r\geq 0,\,\theta\in [0,\pi],\,\varphi\in [0,2\pi].$
The following ansatz for $\Psi_{j,k}$ is used
\begin{equation}\label{alphabeta}
      \psi_{jk} =
      \begin{bmatrix}
       \psi_{jk}^{(1)}\\
       \psi_{jk}^{(2)}\\
       \psi_{jk}^{(3)}\\
       \psi_{jk}^{(4)}
      \end{bmatrix}
      =
      e^{-i\omega t}\frac{\sqrt{T(r)}}{r}  
      \begin{bmatrix}
      \chi^{k}_{j-1/2}\alpha(r) \\
      i\chi^{k}_{j+1/2}\beta(r)
    \end{bmatrix},
\end{equation}
where we restrict the study to solutions with positive parity. Here $\chi^{k}_{j\mp 1/2}(\theta,\varphi)$ is a linear combination of spherical harmonics $Y^{k}_{j\mp 1/2}$ and the basis $e_1 = [1,0]^{\top}$, $e_2 = [0,1]^{\top}$, cf. \cite[Eq. (8)-(9)]{LHHD}. 
A consequence of the restriction to static and spherically symmetric solutions is that the fermion wave functions are separable, i.e. each fermion has the same energy $\omega$ and radial structure and they differ only in their angular dependence. The latter is only visible in the equations through $\kappa$. 
In view of (\ref{alphabeta}) it follows that the matter content of the system is fully encoded in the two scalar functions $\alpha$ and $\beta$. 

We are now in a position to formulate the Einstein-Dirac system in an explicit form. It reads
 \begin{eqnarray}
       \sqrt{A}\alpha'&=& \frac{\kappa}{2r}\alpha - (\omega T + m)\beta, \label{ee1ed}\\
       \sqrt{A} \beta' &=& (\omega T - m)\alpha - \frac{\kappa}{2r}\beta, \label{ee2ed}\\
       rA'&=& 1 - A - 8\pi \kappa\omega T^2(\alpha^2 + \beta^2), \label{ee3ed}\\
       2r A\frac{T'}{T}&= & A-1-8\pi \kappa T\big(\omega T(\alpha^2 + \beta^2) -\frac{\kappa}{r}\alpha\beta - m(\alpha^2 - \beta^2)\big).\label{ee4ed}
  \end{eqnarray}
There is in fact an additional auxiliary Einstein equation, but a solution to the above system satisfies this equation automatically. The last term on the right-hand side of (\ref{ee3ed}) and (\ref{ee4ed}) can be expressed in terms of energy density $\rho$ and radial pressure $p_r$, while the tangential pressure $p_{\perp}$ is present in the auxiliary Einstein equation which is not stated here. These quantities are given by:
     \begin{align*}
         \rho(r) &= \kappa\omega\frac{T^2(r)}{r^2}(\alpha^2(r) + \beta^2(r)),\\
         p_r(r) &= \kappa\frac{T(r)}{r^2}\bigg[\omega T(r)(\alpha^2(r)+\beta^2(r))-m(\alpha^2(r)-\beta^2(r))\\
&\;\;\;\;\; -\kappa\frac{\alpha(r)\beta(r)}{r}\bigg],\\
         p_{\perp}(r) &= \frac{\kappa^2}{2r^{3}}T(r)\alpha(r)\beta(r).
     \end{align*}     

Solutions to the Einstein-Dirac system are found by a numerical procedure involving a scaling of the equations and a method often applied to boundary value problems; a shooting method. 
In \cite{FSY1} the asymptotic flatness condition of the metric fields is replaced by the conditions
\begin{equation}\label{condition_AF}
    \lim_{r\rightarrow\infty} A(r) < \infty, \quad \lim_{r\rightarrow\infty} T(r) < \infty.
\end{equation}
Moreover, the normalization condition imposed on the fermion fields is exchanged by the condition
\begin{equation}\label{condition_normalisation}
    \|\psi_{jk}\|^2 = 4\pi\int_{0}^{\infty} (\alpha(r)^2+\beta(r)^2)\frac{T(r)}{\sqrt{A(r)}}\,\text{d}r < \infty.
\end{equation}
Each of the outlined conditions is involved in the scaling of the Einstein-Dirac equations, in the sense that they define two scaling parameters $\tau$ and $\lambda$,
\begin{align*}
    \tau &:= \lim_{r\rightarrow\infty} T(r),\\
    \lambda &:= \left[ 4\pi\int_{0}^{\infty} (\alpha(r)^2+\beta(r)^2)\frac{T(r)}{\sqrt{A(r)}}\,\text{d}r \right]^{1/2}.
\end{align*}
If a solution $(\alpha,\beta,T,A)$ is given which satisfies the Einstein equations together with conditions (\ref{condition_AF}) and (\ref{condition_normalisation}), then the functions

\begin{align*}
   &\Tilde{A}(r) := A(\lambda r), \quad \Tilde{T}(r) := \frac{1}{\tau}T(\lambda r),\\
   &\Tilde{\alpha}(r) := \sqrt{\frac{\tau}{\lambda}}\alpha(\lambda r), \quad \Tilde{\beta}(r) := \sqrt{\frac{\tau}{\lambda}}\beta(\lambda r),
\end{align*}
satisfy the Einstein equations with $m$ and $\omega$ replaced by 
\begin{equation*}
    \Tilde{m}=\lambda m, \quad \Tilde{\omega}=\lambda \omega \tau.
\end{equation*}
The solution $(\Tilde{\alpha},\Tilde{\beta},\Tilde{A},\Tilde{T})$ is asymptotically flat and the original normalization condition of the fermion fields is satisfied. 
The scaled parameters $\tilde{m}$ and $\tilde{\omega}$ are obtained from the shooting method. A shooting method is an iterative procedure in which a parameter is changed until the solutions to a differential equation satisfy the boundary conditions. 
In the case of the Einstein-Dirac equations, the parameter $\omega$, is the parameter that is changed during each step, and it is varied until the asymptotic properties described above are fulfilled for both the metric and fermion fields. The regularity condition in the centre, $A(0)=1$, is implicitly imposed by integrating the equations for a small initial radius $r_0>0$. 
Solutions for one iterate of the parameter were generated by a Taylor series ODE-solver, which was built with arbitrary floating point precision in mind; the function is available in the Python library: \texttt{mpmath} \cite{mpmath}. 
High precision is crucial in order to generate solutions for increasing $\kappa$. The reason is that a too small parameter yields divergent solutions while a too large parameter yields solutions with an oscillating behaviour.

Since we restrict this study to ground states,
we look for the smallest value of $\omega$, which yields normalizable solutions of $\psi_{jk}$. 
The high level of precision required an efficient update rule for the search parameter $\omega$. These requirements were met by the classical binary search algorithm, which acts on an input interval $[\omega_{\text{low}},\omega_{\text{high}}]$.
All such search intervals were detected using visual inspection. If physically significant solutions are found and scaled as outlined above, the central redshift is directly computed from the scaled metric field $\Tilde{T}$,
\begin{equation}
    \label{eq:redshift_z}
    z:=\Tilde{T}(0)-1\approx \Tilde{T}(r_0)-1.
\end{equation}
The central redshift of the solution is implicitly decided by a parameter, $\alpha_1>0$, used to define the initial conditions to the Einstein-Dirac equations, which were defined by a Taylor series around $r=0$.
The expressions for the small-radius asymptotic expansion are given in \cite{LHHD} and read
\begin{subequations}
\begingroup{}{}
\begin{alignat}{1}
    \alpha(r) &= \alpha_1r^{\kappa/2} + \mathcal{O}(r^{(\kappa/2 + 2)}), \\
         \beta(r) &= \frac{1}{\kappa + 1}(\omega T_0 -m)\alpha_1 r^{(\kappa/2 + 1)}  + \mathcal{O}(r^{(\kappa/2 + 3)}), \\
         A(r) &= 1 - 8\pi \omega T_0^2 \alpha_{1}^2\frac{\kappa}{\kappa + 1}r^{\kappa} +  \mathcal{O}(r^{(\kappa + 2)}), \\
         T(r) &= T_0 - 4\pi T_{0}^2\alpha_{1}^2 \frac{1}{\kappa + 1}(2\omega T_0 - m)r^{\kappa} +  \mathcal{O}(r^{(\kappa + 2)}).
\end{alignat}
\label{eq:IV_ED}
\endgroup
\end{subequations}
For all solutions presented in this work $T_0=1$ and $m=1$. 

\section{The static Einstein\,-Vlasov system}\label{sec-EV}
The solutions of the Einstein\,-Vlasov system that we compare with the solutions of the Einstein-Dirac system are obtained by using the numerical strategy described in \cite{AR}. We therefore follow the notation in \cite{AR} and write the metric as 
\begin{equation}\label{metric-EV}
ds^{2}=-e^{2\mu(r)}dt^{2}+e^{2\lambda(r)}dr^{2}
+r^{2}(d\theta^{2}+\sin^{2}{\theta}d\varphi^{2}),
\end{equation}
where $r\geq 0,\,\theta\in [0,\pi],\,\varphi\in [0,2\pi].$ Clearly, the metric (\ref{metric-EV}) has an identical form as the metric (\ref{metric-ED}) in the previous section, only the notation differs. 
Asymptotic flatness is expressed by the boundary conditions
\begin{displaymath}
\lim_{r\rightarrow\infty}\lambda(r)=\lim_{r\rightarrow\infty}\mu(r)
=0. 
\end{displaymath}
We now formulate the static Einstein\,-Vlasov system; for an introduction to the Einstein\,-Vlasov system we refer to \cite{AGS,A3}. 
The static Einstein\,-Vlasov system is given by the Einstein equations 
\begin{eqnarray}
&\displaystyle e^{-2\lambda}(2r\lambda_{r}-1)+1=8\pi r^2\rho,&\label{ee12}\\
&\displaystyle e^{-2\lambda}(2r\mu_{r}+1)-1=8\pi r^2
p,&\label{ee22}\\
&\displaystyle
\mu_{rr}+(\mu_{r}-\lambda_{r})\left(\mu_{r}+\frac{1}{r}\right)= 8\pi
p_T e^{2\lambda},&\label{ee4}
\end{eqnarray}
together with the static Vlasov equation 
\begin{equation}
\frac{w}{\varepsilon}\partial_{r}f -\left(\mu_{r} \varepsilon-
\frac{L}{r^3\varepsilon}\right)\partial_{w}f=0,\label{vlasov}
\end{equation}
where
\begin{equation*}
\varepsilon=\varepsilon(r,w,L)=\sqrt{1+w^{2}+L/r^{2}}.
\end{equation*}
The matter quantities are defined by
\begin{eqnarray}
\rho(r)&=&\frac{\pi}{r^{2}}
\int_{-\infty}^{\infty}\int_{0}^{\infty}\varepsilon(r,w,L) f(r,w,L)\;dLdw,\label{rho21}\\
p(r)&=&\frac{\pi}{r^{2}}\int_{-\infty}^{\infty}\int_{0}^{\infty}
\frac{w^{2}}{\varepsilon(r,w,L)}f(r,w,L)\;dLdw,\label{p21}\\
p_T(r)&=&\frac{\pi}{2r^{4}}\int_{-\infty}^{\infty}\int_{0}^{\infty}\frac{L}{\varepsilon(r,w,L)}f(r,w,L)\;
dLdw. 
\end{eqnarray}
The variables $w$ and $L$ can be thought of as the momentum in the radial direction and the square of the angular momentum, respectively.

The matter quantities $\rho, p$ and $p_T$ are the energy density, the radial pressure, and the tangential pressure, respectively. 
The system of equations above are not independent, and we study the reduced system (\ref{ee12})-(\ref{ee22}) together with (\ref{vlasov}) and (\ref{rho21})-(\ref{p21}). 
It is straightforward to show that a solution to the reduced system is a solution to the full system. 

Define $$E=e^{\mu}\varepsilon,$$ then the ansatz 
\begin{equation}
f(r,w,L)=\Phi(E,L), \label{ansatz}
\end{equation}
where $\Phi$ is a given function, satisfies (\ref{vlasov}). 
By inserting this ansatz into (\ref{rho21})-(\ref{p21}) the system of equations reduces to a system where the metric coefficients $\mu$ and $\lambda$ alone are the unknowns. 
This has turned out to be an efficient method for constructing static solutions, and we will use this approach here. 
The following form of $\Phi$ will be used 
\begin{equation}
\Phi(E,L)=(E_0-E)^k_{+}(L-L_0)_{+}^l,\label{pol}
\end{equation}
where $l\geq 1/2,\,k\geq 0,\,L_0>0,\; E_0>0,$ and $x_{+}:=\max\{x,0\}$.
In the Newtonian case with $l=L_0=0,$ this ansatz leads
to steady states with a polytropic equation of state, and we refer to (\ref{pol}) as the polytropic ansatz. 

In the next section, a few static solutions of the Einstein\,-Vlasov system are displayed where the ansatz (\ref{pol}) has been used. These solutions were obtained using the numerical scheme outlined in \cite{AR}. In this work the notation
\begin{equation}\label{y0}
y(r):=\frac{e^{\mu(r)}}{E_0}
\end{equation}
is used. The central redshift $z$ of the static solution is determined by $y_0:=y(0)$ from the formula
\[
z=\frac{y(R)}{y_0}-1,
\]
where $R$ is the outer boundary of the solution. 
Consequently, the parameters that determine a static solution of the Einstein-Vlasov system with a polytropic ansatz are $y_0,k,l$ and $L_0$, cf. \cite{AR}. 

\section{Solution properties with respect to $\kappa$ and $z$}\label{sec-classify}
\begin{figure}[htb]
    \centering
    \begin{tikzpicture}[
        image/.style = {text width=0.45\textwidth, 
                        inner sep=0pt, outer sep=0pt},
        eqn/.style = {anchor=south, text=black, font=\small},
        node distance = 1mm and 1mm
    ]

\node [image] (frame1)
    {\includegraphics[width=\linewidth]{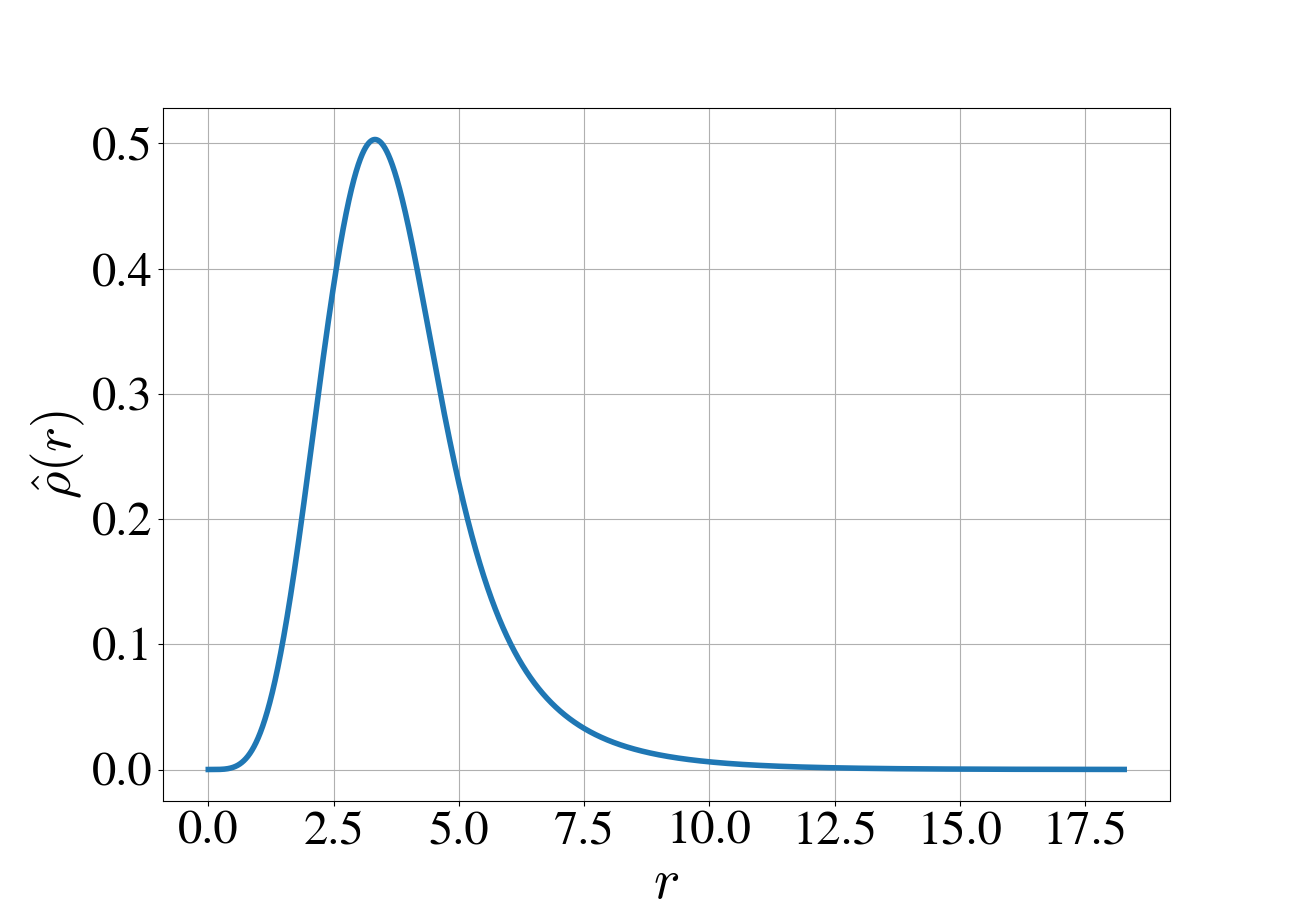}};
    \node[eqn] at ([yshift=-6mm]frame1.north) {$\kappa = 4,\,\,z=2.2$};

\node [image,right=of frame1] (frame2) 
    {\includegraphics[width=\linewidth]{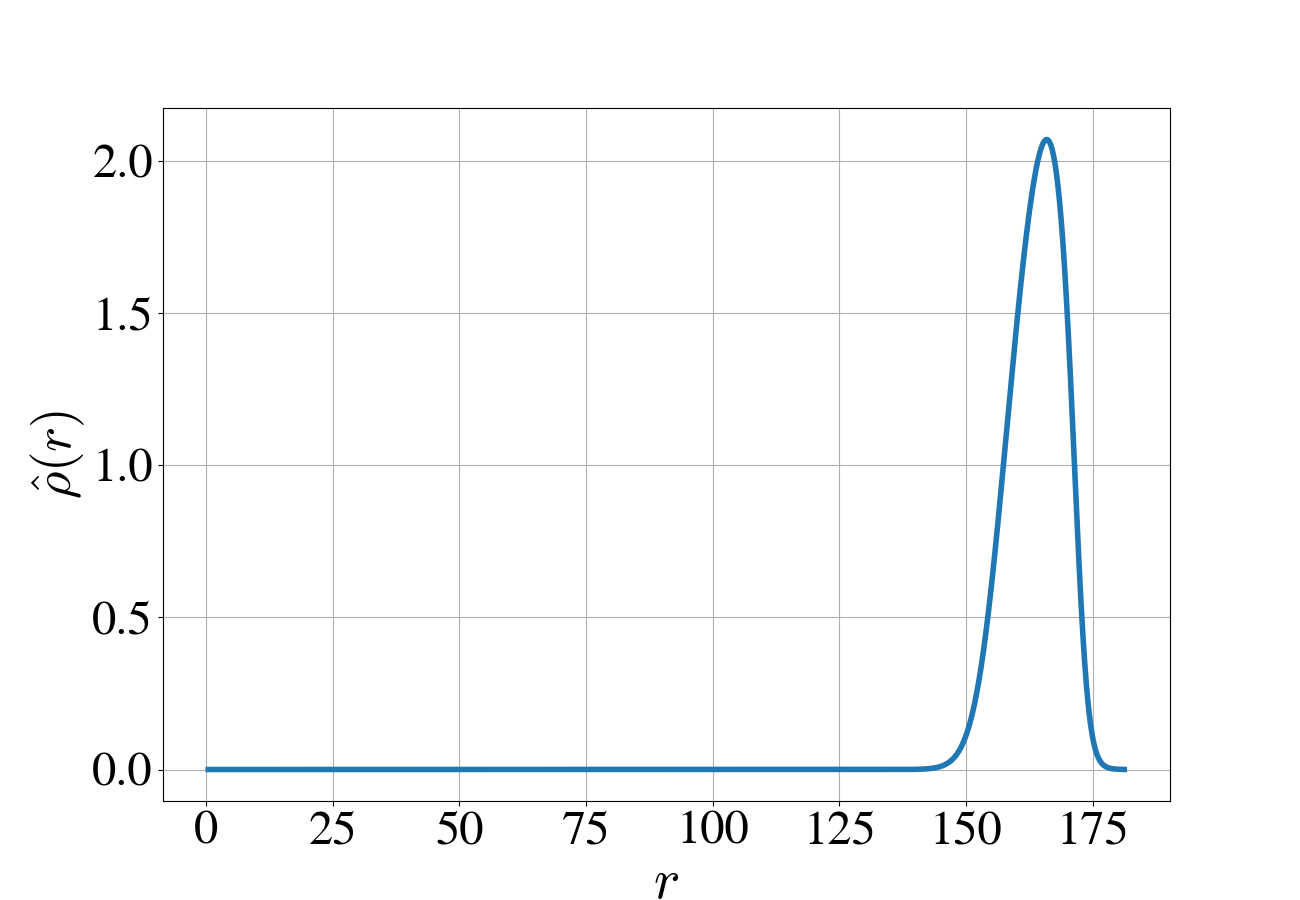}};
    \node[eqn] at ([yshift=-6mm]frame2.north) {$\kappa = 180,\,\,z=1.8$};
\end{tikzpicture}
    \caption{To the left a solution for $\kappa=4$ and to the right a solution for $\kappa=180$. In both cases the central redshift $z\approx 2$ which is comparatively small and only single-peak solutions arise.}\label{fig:kz1}
\end{figure}

As outlined in Section \ref{sec-ED} there is a continuous family of solutions to the Einstein-Dirac system for a fixed $\kappa$ parametrized by the central redshift $z$. As $z$ increases, the central region of the solution becomes more compressed. For small $z$, the energy density of the solution exhibits a single peak independently of the value of $\kappa$, as seen in Figure \ref{fig:kz1} where $z\approx 2$ in the two cases $\kappa=4$ and $\kappa=180$. 

\begin{figure}[htbp]
\begin{center}
\scalebox{.28}
{\includegraphics{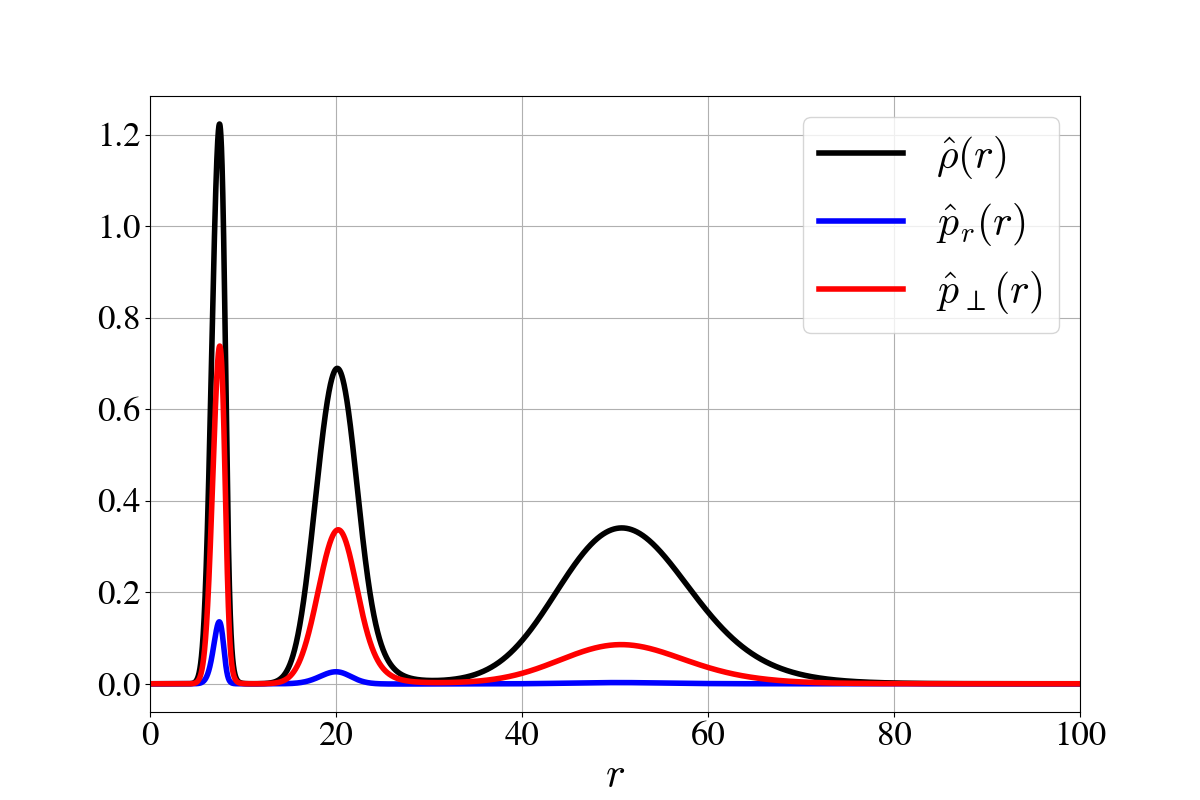}}
\end{center}
\caption{The energy density, the radial pressure and the tangential pressure are shown in the case $\kappa=32$ and $z=7$. 
The tangential pressure dominates the radial pressure which is also the case for the solutions of the Einstein-Vlasov system that we consider in this work.}\label{fig:all_T}
\end{figure}

\begin{remark}
In several figures we show the energy density of a solution and we often use a logarithmic scale since for multi-peak solutions the amplitude of the inner peak is much larger than for the other peaks. We therefore introduce the notation
\[
\hat{\rho}=\log(4\pi r^2\rho+1).
\]
We define $\hat{p}$ analogously; for the solutions we have considered in this work $4\pi r^2p+1>0$. 
The energy density is a macroscopic quantity which is natural to consider when we compare solutions of the Einstein-Dirac system and the Einstein\,-Vlasov system. We could also show the pressure components, but their structure is similar, and we typically only show the energy density; however, see Figure \ref{fig:all_T} where the energy density, the radial pressure, and the tangential pressure are shown in the same figure. Other natural quantities to consider are the fermion fields $\alpha$ and $\beta$, as well as the metric fields. We refer to \cite{LHHD} and \cite{FSY1} where these quantities are depicted rather than the components of the energy-momentum tensor. 
\end{remark}

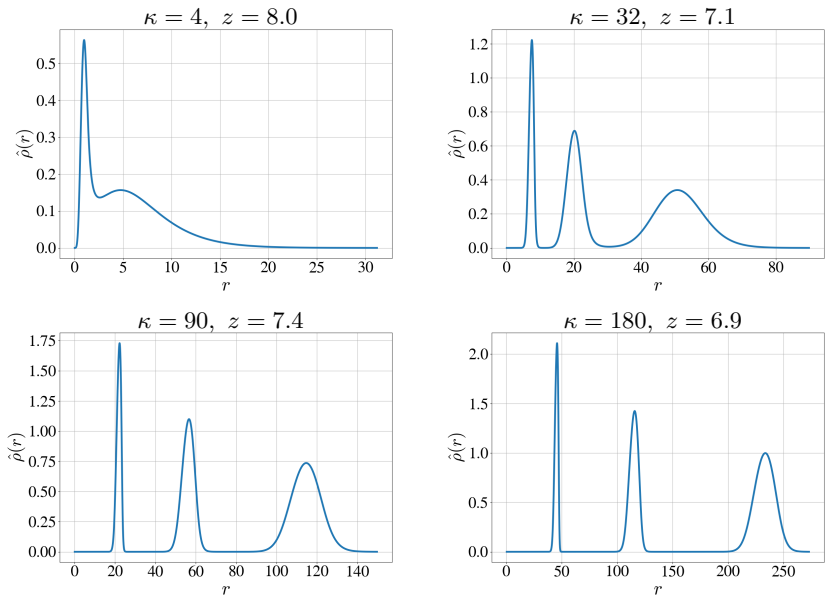
\begin{figure}[htb]
    \centering
    \begin{tikzpicture}[
        image/.style = {text width=0.45\textwidth, 
                        inner sep=0pt, outer sep=0pt},
        eqn/.style = {anchor=south, text=black, font=\footnotesize},
        node distance = 1mm and 0.5mm
    ]  
\node [image] (frame1)
    {\includegraphics[width=\linewidth]{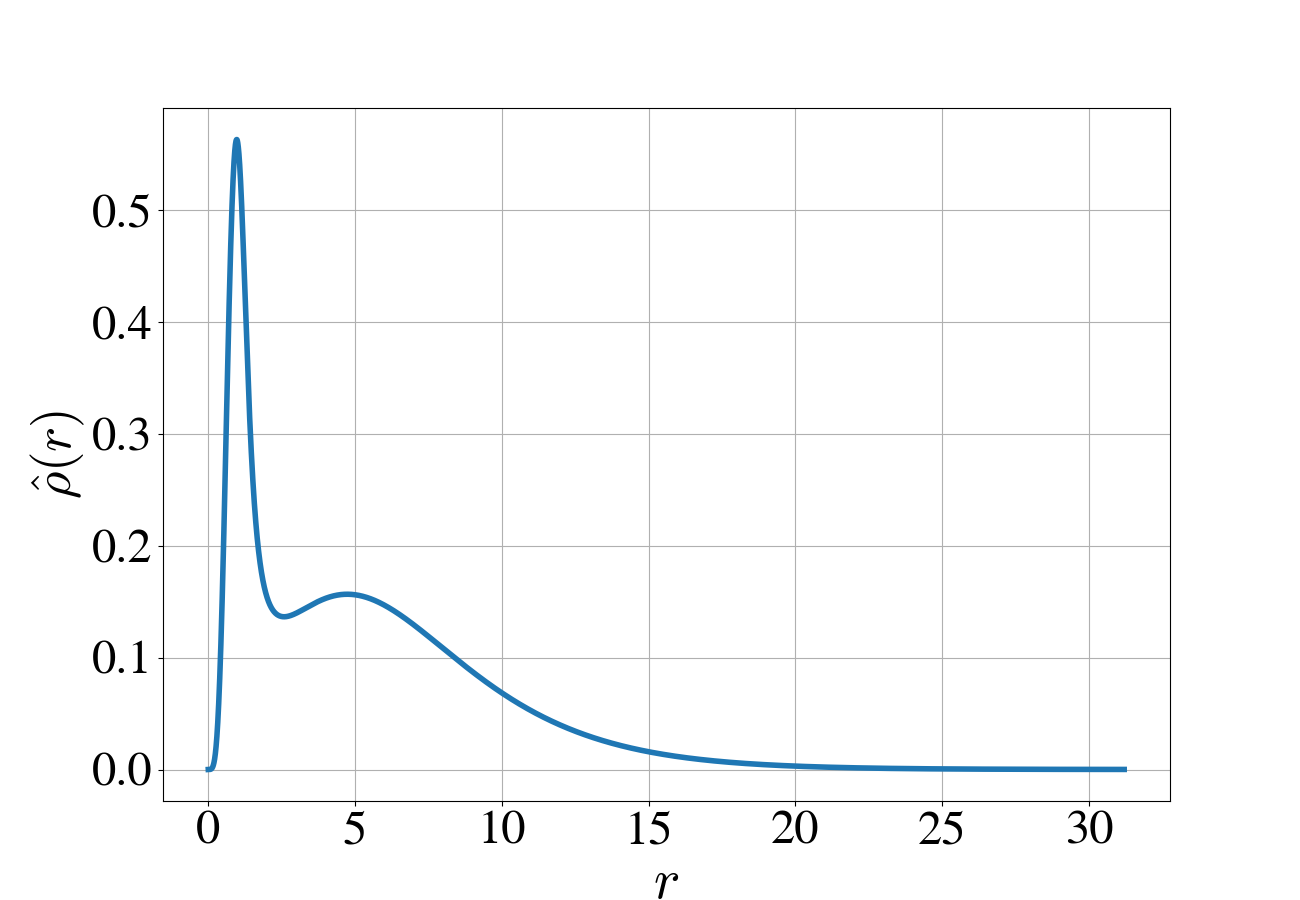}};
    \node[eqn] at ([yshift=-6mm]frame1.north) {$\kappa = 4,\,\,z=8.0$};
\node [image,right=of frame1] (frame2) 
    {\includegraphics[width=\linewidth]{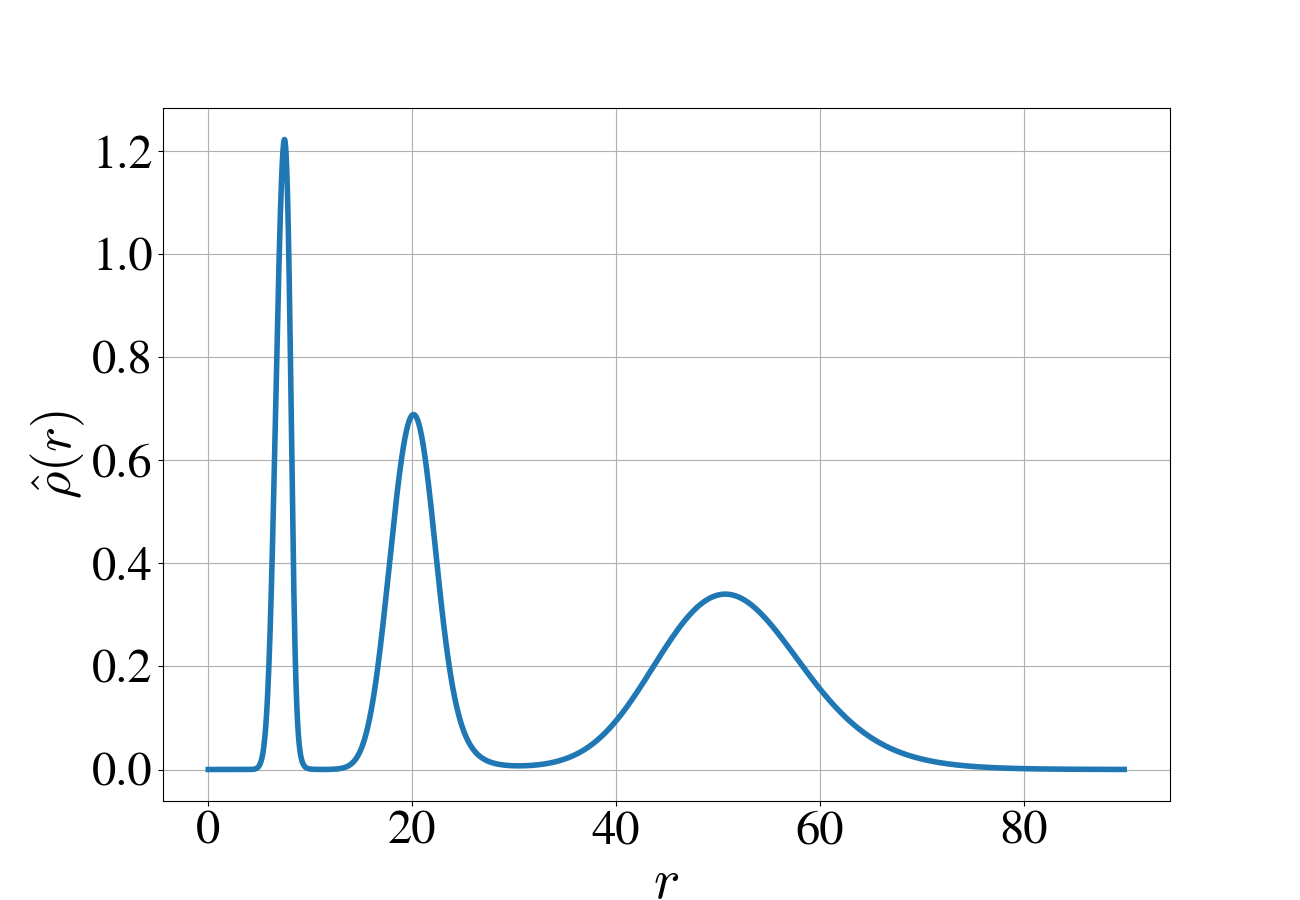}};
    \node[eqn] at ([yshift=-6mm]frame2.north) {$\kappa = 32,\,\,z=7.1$};
\node[image,below=of frame1] (frame3)
    {\includegraphics[width=\linewidth]{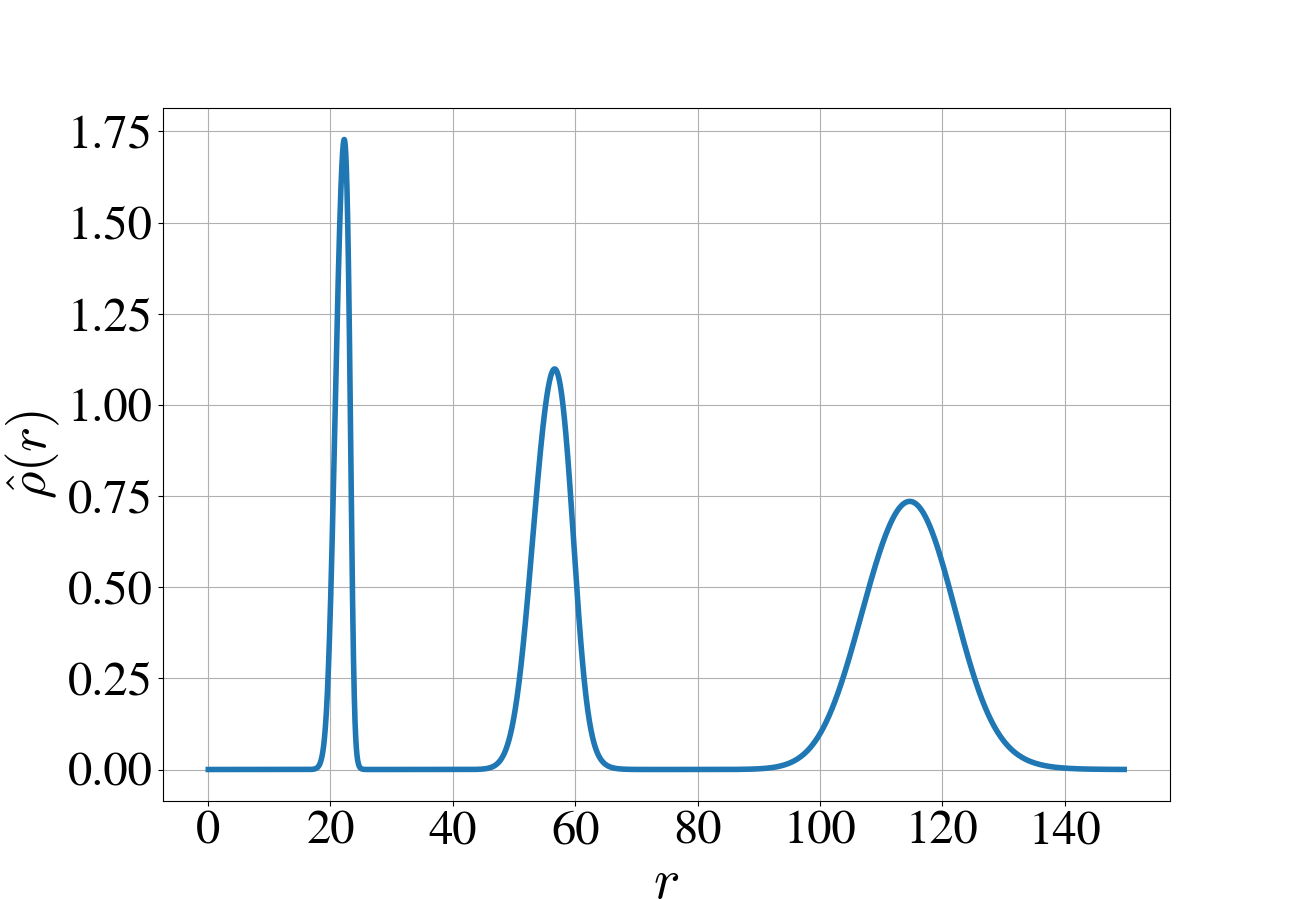}};
    \node[eqn] at ([yshift=-6mm]frame3.north) {$\kappa = 90,\,\,z=7.4$};
\node[image,right=of frame3] (frame4)
    {\includegraphics[width=\linewidth]{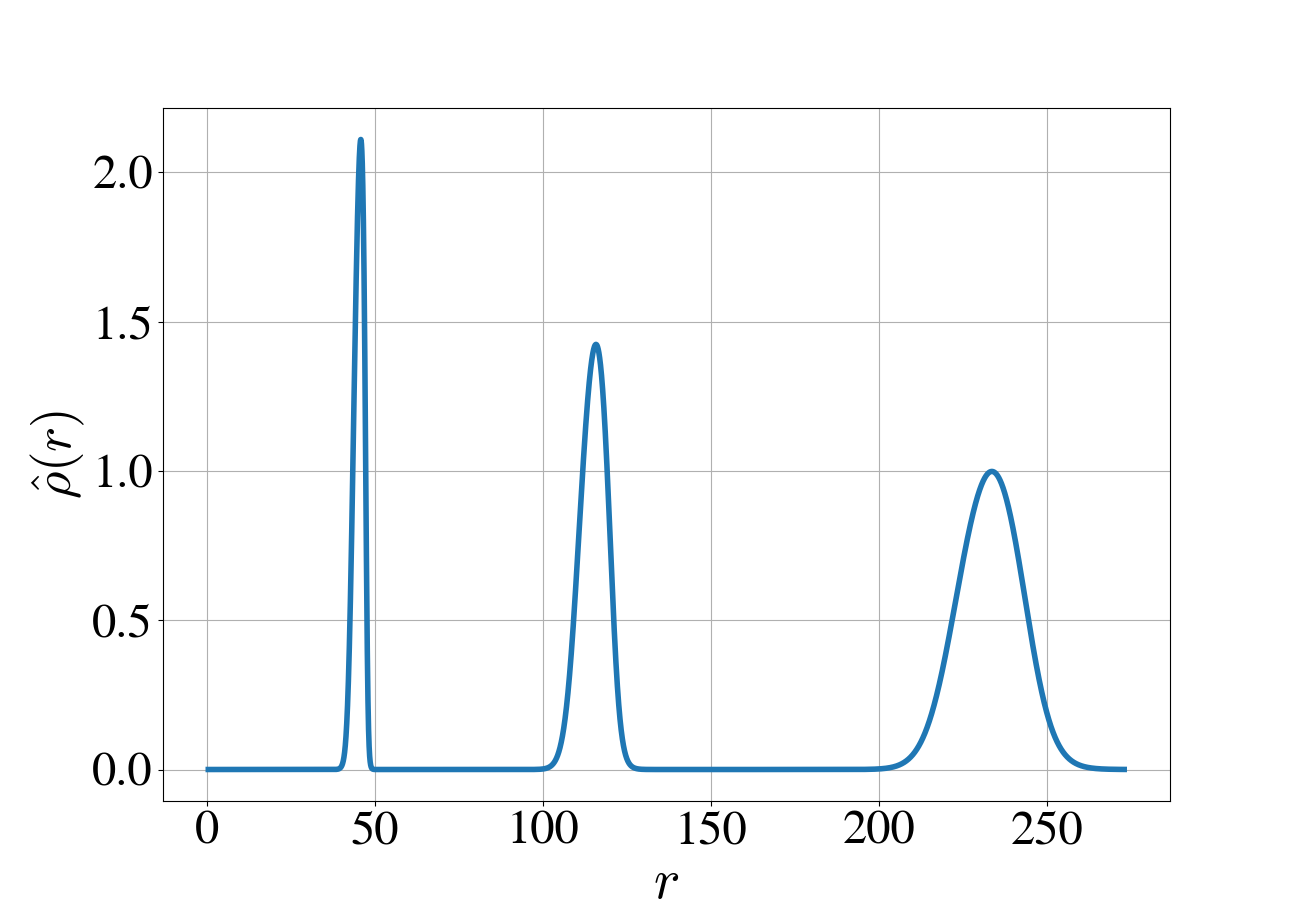}};
    \node[eqn] at ([yshift=-6mm]frame4.north) {$\kappa = 180,\,\,z=6.9$};
\end{tikzpicture}
    \caption{A sequence of solutions corresponding to $\kappa=4,32,90,180$, with a similar central redshift, is displayed. The redshift $z\approx 7.5$ is sufficiently large in order to obtain solutions with a multi-peak strucure.}\label{fig:kz2}
\end{figure}

As $z$ increases, the multi-peak structure of the solutions starts to appear, which is shown in Figure \ref{fig:kz2} where $z\approx 7$. We find that the separation of the peaks becomes more pronounced as $\kappa$ increases, 
and it is evident that the number of fermions is essential for a distinct multi-peak structure. The main focus of this work is on cases where $\kappa$ and $z$ are comparatively large, which yields highly compact configurations as will be seen below.

\begin{figure}[htb]
    \centering
    \begin{tikzpicture}[
        image/.style = {text width=0.45\textwidth, 
                        inner sep=0pt, outer sep=0pt},
        eqn/.style = {anchor=south, text=black, font=\footnotesize},
        node distance = 1mm and 0.5mm
    ] 
\node [image] (frame1)
    {\includegraphics[width=\linewidth]{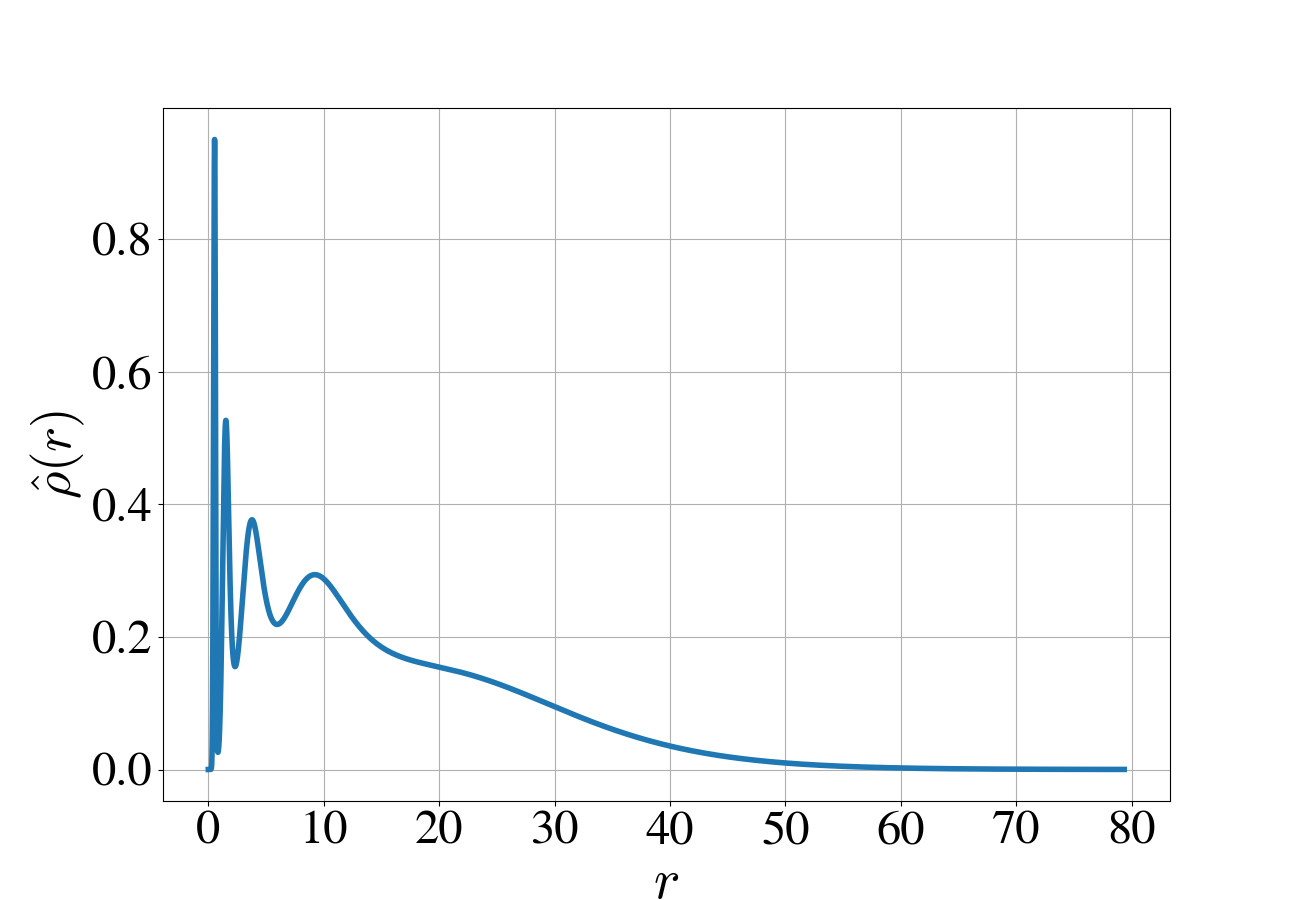}};
    \node[eqn] at ([yshift=-6mm]frame1.north) {$\kappa = 16,\,\,z=59$};
\node [image,right=of frame1] (frame2) 
    {\includegraphics[width=\linewidth]{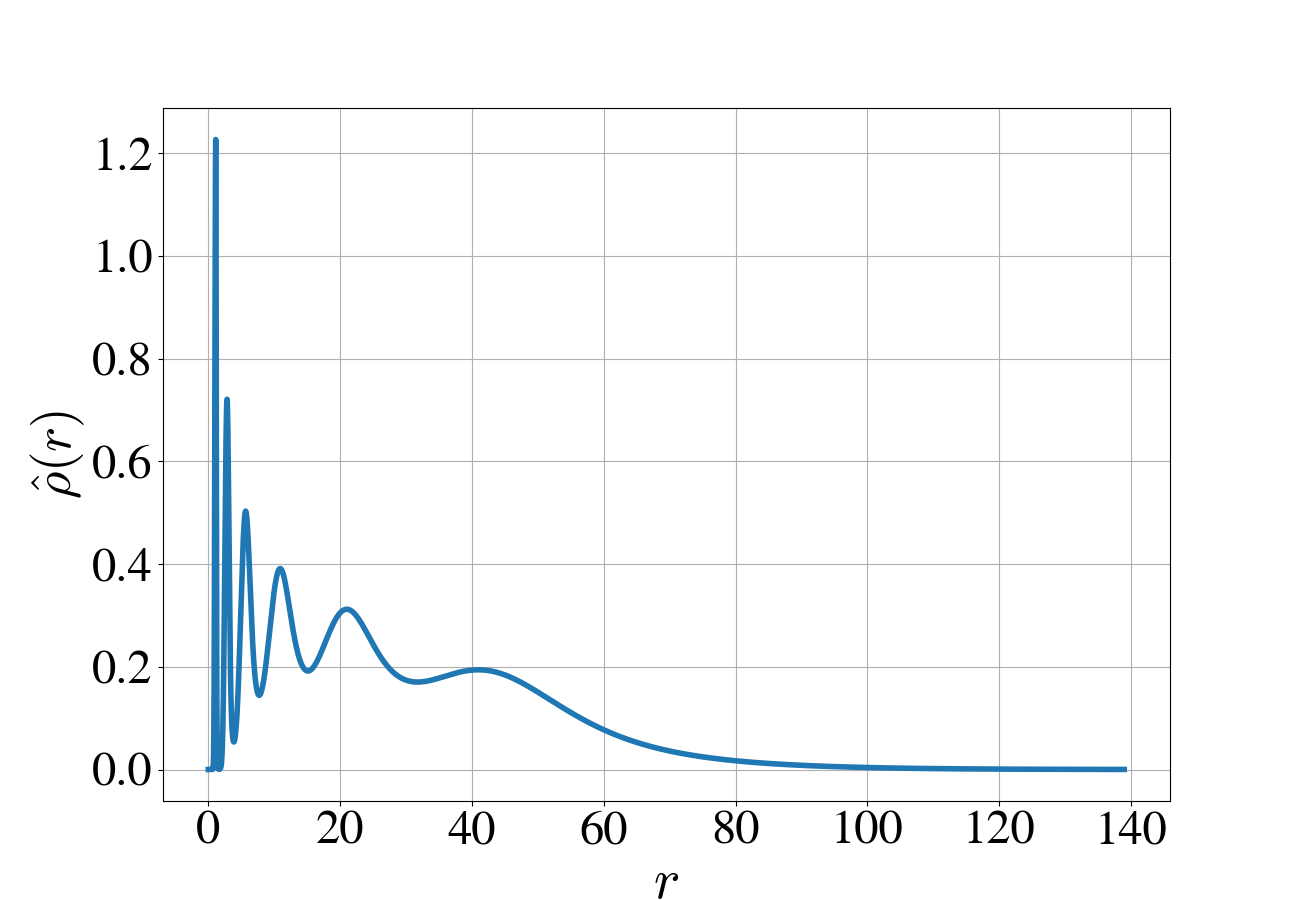}};
    \node[eqn] at ([yshift=-6mm]frame2.north) {$\kappa = 32,\,\,z=56$};
\node[image,below=of frame1] (frame3)
    {\includegraphics[width=\linewidth]{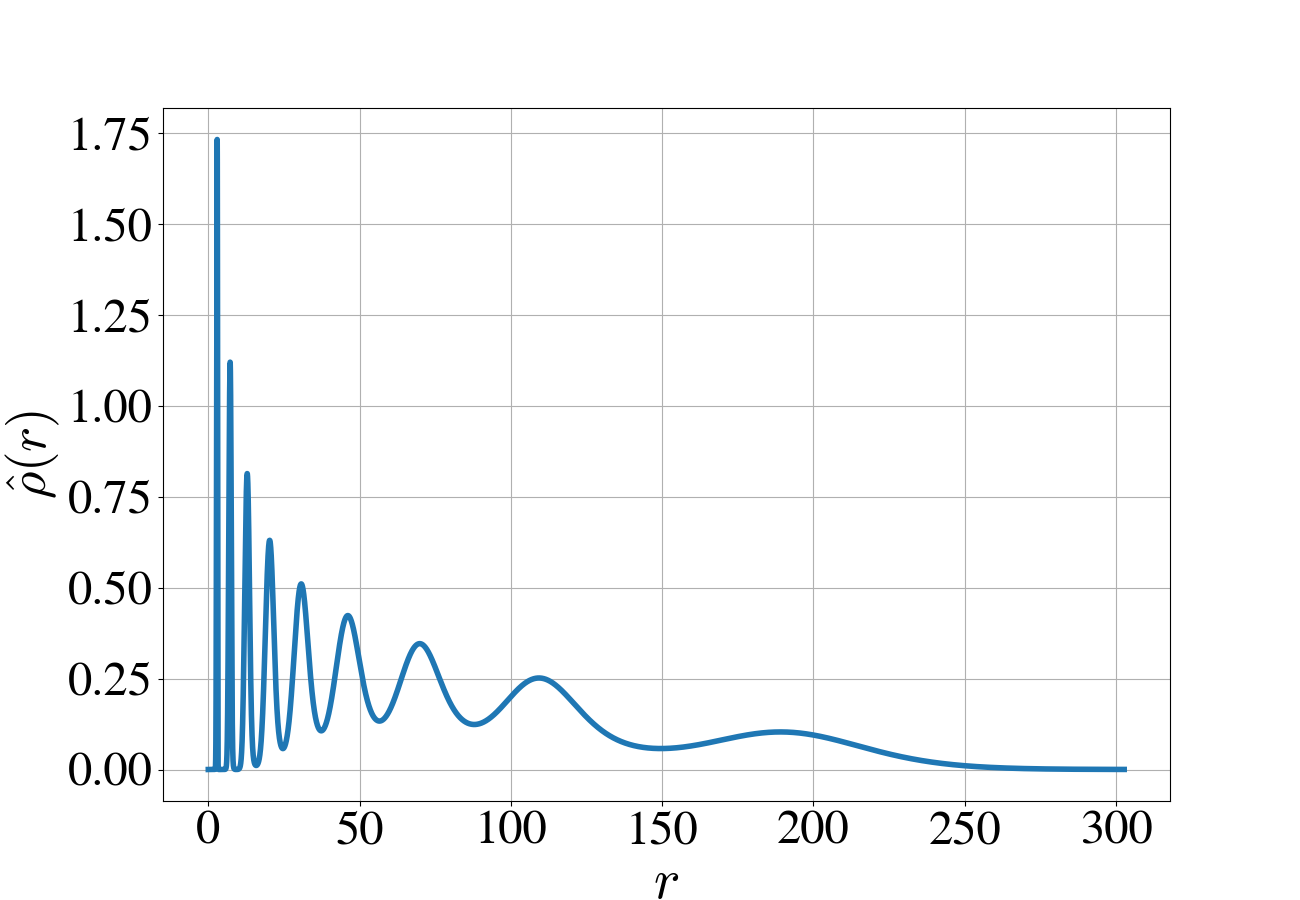}};
    \node[eqn] at ([yshift=-6mm]frame3.north) {$\kappa = 90,\,\,z=55$};
\node[image,right=of frame3] (frame4)
    {\includegraphics[width=\linewidth]{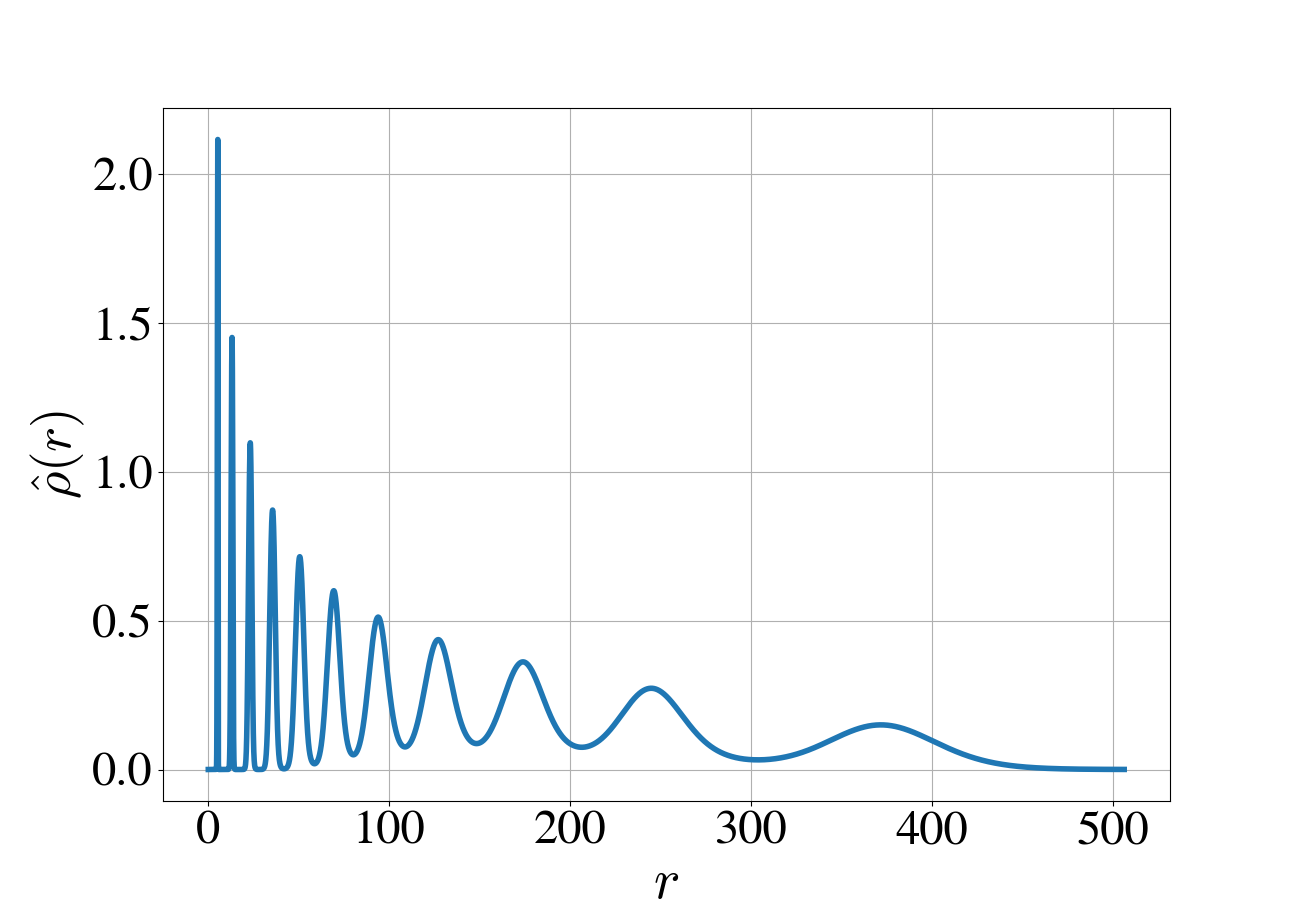}};
    \node[eqn] at ([yshift=-6mm]frame4.north) {$\kappa = 180,\,\,z=57$};
\end{tikzpicture}
     \caption{A sequence of solutions corresponding to $\kappa=16,32,90,180$, with a similar redshift $z\approx 57$, is shown. The number of peaks increases as $\kappa$ increases and the peaks also become more narrow.}\label{fig:kz3}
\end{figure}

Taking even larger values of $z$, the number of peaks increases as shown in Figure \ref{fig:kz3} where $z\approx 57$ and $\kappa=16,32,90,180$. Furthermore, the width of the peaks shrinks as $\kappa$ increases (see Section \ref{sec-sharpness} for a detailed discussion of this issue). A natural question to ask is what happens in the limit $\kappa\to\infty$. It is suggested in \cite{LHHD} that the fermion wave function will split into a series of delta functions in this limit, but it is an open question.

In the case of the Einstein\,-Vlasov system there is no dependence on $\kappa$ but as described in \cite{AR} the solutions depend in a similar way with respect to the central redshift $z$, that is, for small $z$ the solutions are single-peaked and as $z$ increases the multi-peak structure starts to appear. 
Moreover, as $z$ increases, not only are the solutions to the two systems multi-peaked, but the solutions also become more compact, i.e. $\Gamma$ increases. 
We refer to "highly compact solutions" when $\Gamma$ is comparatively large. In \cite{LHHD} the corresponding terminology is referred to as the phenomenon of self-trapping. 
In the following sections, we will look closer to the remarkable similarity between highly compact solutions, or solutions which show self-trapping behaviour, of these two systems.

\section{Striking similarity of compact solutions}\label{sec-comparison}
In the previous section, some properties of solutions to the Einstein-Dirac system were briefly investigated with respect to $\kappa$ and $z$. The features of static solutions, in particular shell solutions, to the Einstein\,-Vlasov system were numerically explored in \cite{AR} for the polytropic ansatz (\ref{pol}) and a similar structure was found. The energy density possesses many peaks if the central redshift $z$ is sufficiently high. The different peaks are separated by vacuum regions or regions where the energy density is very low. 
The structure of the solutions to the Einstein-Dirac system is clearly very similar, but it should be noticed that the energy density of the separating regions can never vanish completely in that case.

\begin{figure}[htbp]
\begin{center}
\hspace*{-6mm}
\scalebox{.30}{\includegraphics{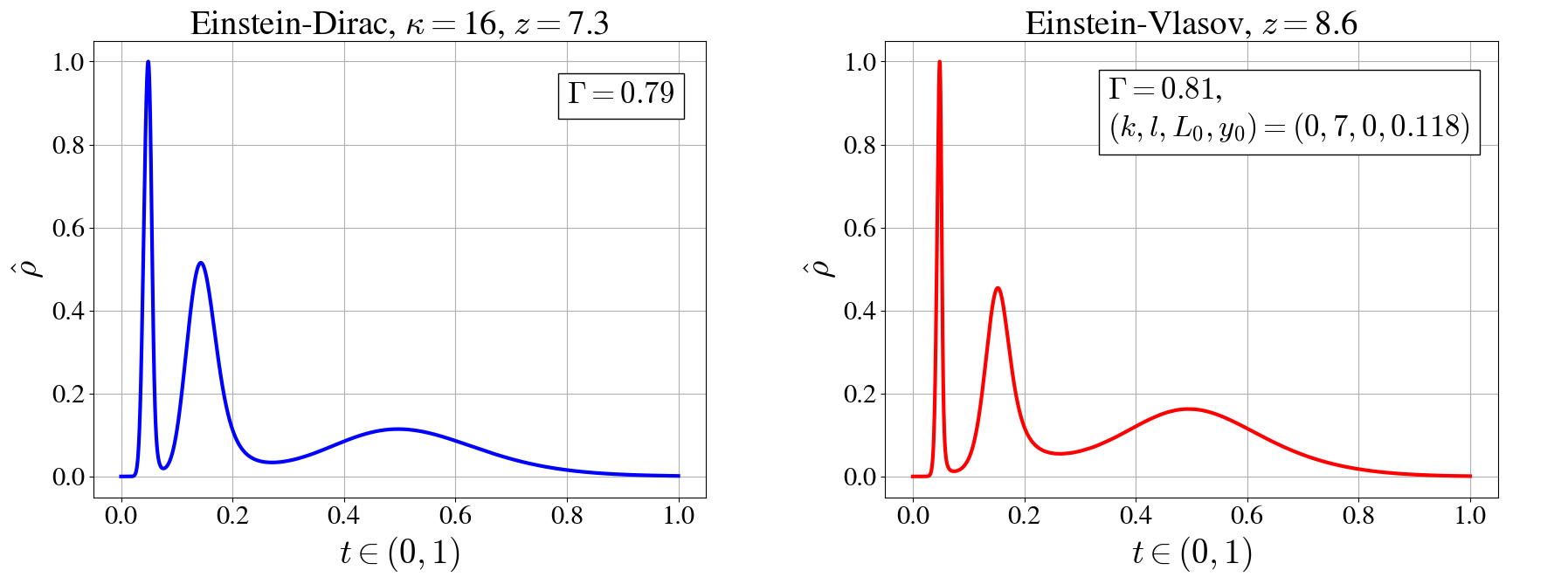}}
\end{center}
\caption{A solution of the Einstein-Dirac system for $\kappa=16$ is shown to the left and an analogous solution to the Einstein\,-Vlasov system is shown to the right. The compactness ratio $\Gamma$ is similar for both solutions. }\label{fig:comparison-1}
\end{figure}

The purpose of the present section is to study how closely multi-peak solutions of the two systems agree. In principle, it could be that the peaks have very different relative amplitudes or that the spacing between the peaks is very different. However, we find that for $\kappa$ comparatively large, solutions of the Einstein\,-Vlasov system of polytropic type, as described in Section \ref{sec-EV}, can be constructed that agree with remarkable precision to solutions of the Einstein-Dirac system. In the following, we give two examples in the cases $\kappa=16$ and $\kappa=90$. In the left sub-figure in Figure \ref{fig:comparison-1} the energy density is shown
for a solution to the Einstein-Dirac system with $\kappa=16$ and $z=16.9$. The maximum compactness of this solution is $\Gamma=0.79$. In the right sub-figure of Figure \ref{fig:comparison-1} a solution to the Einstein\,-Vlasov system with $\Gamma=0.81$ is depicted. Clearly, the similarity between these solutions is striking. 

\begin{figure}[htbp]
\begin{center}
\hspace*{-7mm}
\scalebox{.30}{\includegraphics{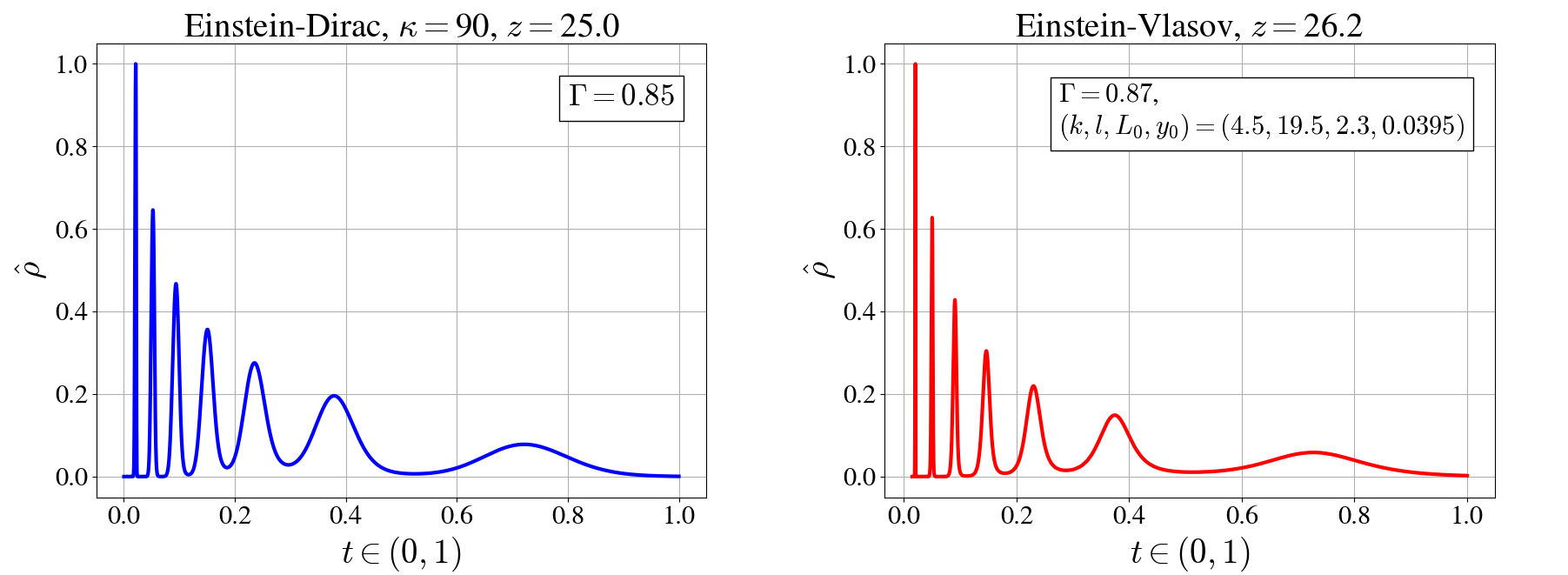}}
\end{center}
\caption{Another example of notably similar solutions of the Einstein-Dirac system and the Einstein\,-Vlasov system. In this case $\kappa=90$.}\label{fig:comparison-2}
\end{figure}

In Figure \ref{fig:comparison-2} we show an example where $\kappa=90$. Again, the structure of the energy density and the compactness of the solutions are notably similar; $\Gamma=0.85$ and $\Gamma=0.87$, respectively. It is possible to find many more examples of this kind, but here we restrict ourselves to these two examples. We emphasize that such a close correspondence between solutions we do not find if $\kappa$ is small, e.g., when $\kappa=2$ or $\kappa=4$, as will be clear in Section \ref{sec-pressure}. It is an interesting analytic problem to try to understand the relation between the structure of the solutions of these systems as $\kappa\to \infty$. 

Let us finish this section with a discussion about multi-peak solutions beyond spherical symmetry. Numerical simulations indicate that the spherically symmetric multi-peak solutions are unstable for both systems, cf. \cite{AR2} and \cite{LHHD}. Together with Ames and Logg, the first author constructed axially symmetric stationary solutions of the Einstein\,-Vlasov system in \cite{AAL1,AAL2}. In the course of these works the authors tried to find axially symmetric multi-peak solutions but failed. It should be pointed out that the numerical method in this case is very different, since it is based on an iteration algorithm rather than solving an ordinary integro-differential equation as in \cite{AR}; see \cite{AA,AKR} for a discussion about the iterative algorithm. 
Now, there are arguments that the stability properties in the axially symmetric case can be quite different compared to the spherically symmetric case if the total angular momentum is non-zero, in particular if it is large compared to its mass; cf. \cite{AA,AAL2}. Since the multi-peak structure is present in the spherically symmetric case for both systems, we find the question whether or not this property is a generic feature of these systems important. It would be very interesting if multi-peak axisymmetric solutions could be constructed with non-vanishing angular momentum for any of these systems. We note here that in the case of boson stars Herdeiro et al. \cite{LHR} in fact have constructed axisymmetric multi-peaked (ring-like) solutions.

\section{Regions of negative pressure and the transition to classical behaviour}\label{sec-pressure}

\begin{figure}[htb]
    \centering
    \begin{tikzpicture}[
        image/.style = {text width=0.45\textwidth, 
                        inner sep=0pt, outer sep=0pt},
        eqn/.style = {anchor=south, text=black, font=\footnotesize},
        node distance = 1mm and 1mm]  
\node [image] (frame1)
    {\includegraphics[width=\linewidth]{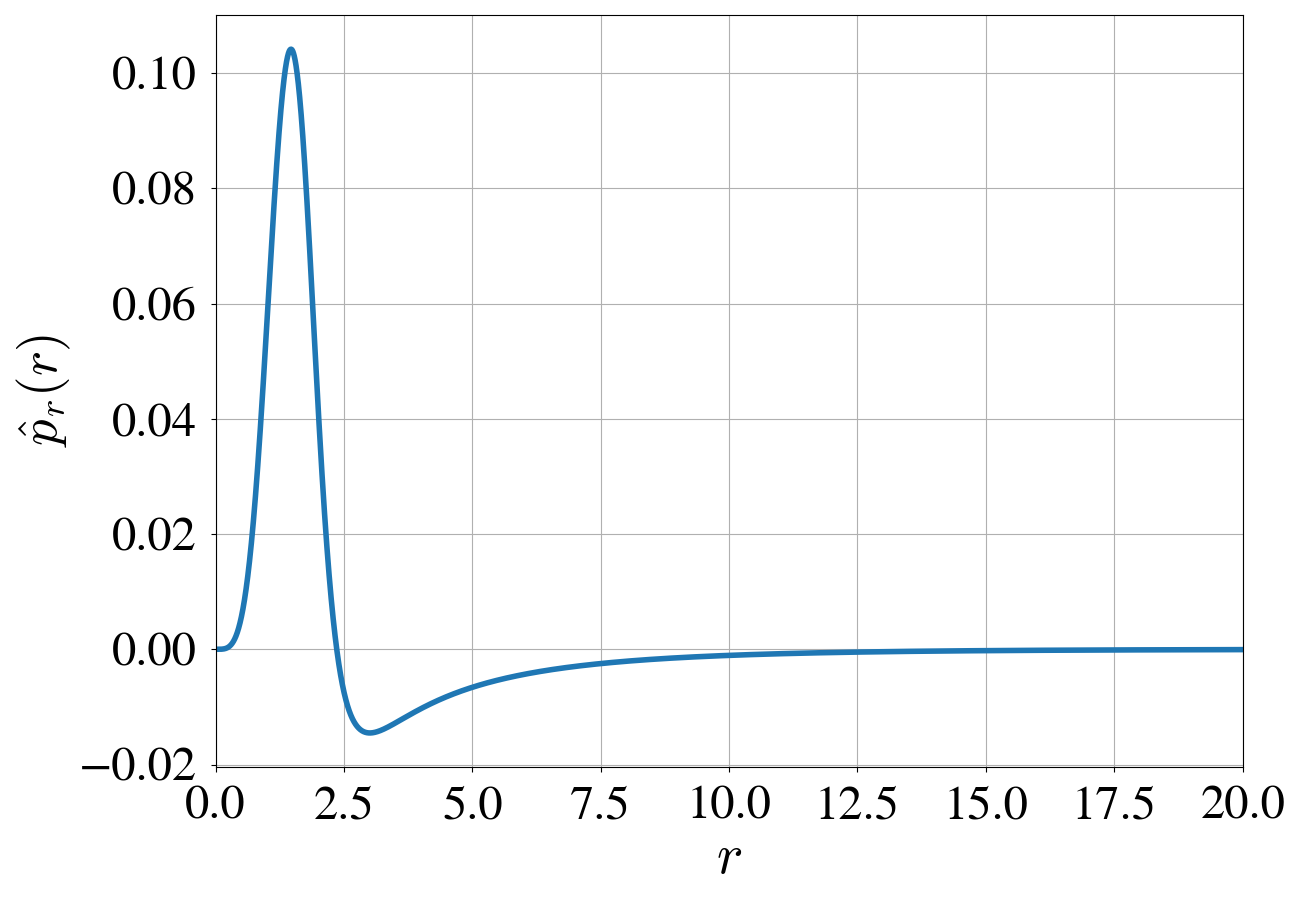}};
    \node[eqn] at ([yshift=-2mm]frame1.north) {$\kappa = 4,\,\,z=4.9$};
\node [image,right=of frame1] (frame2) 
    {\includegraphics[width=\linewidth]{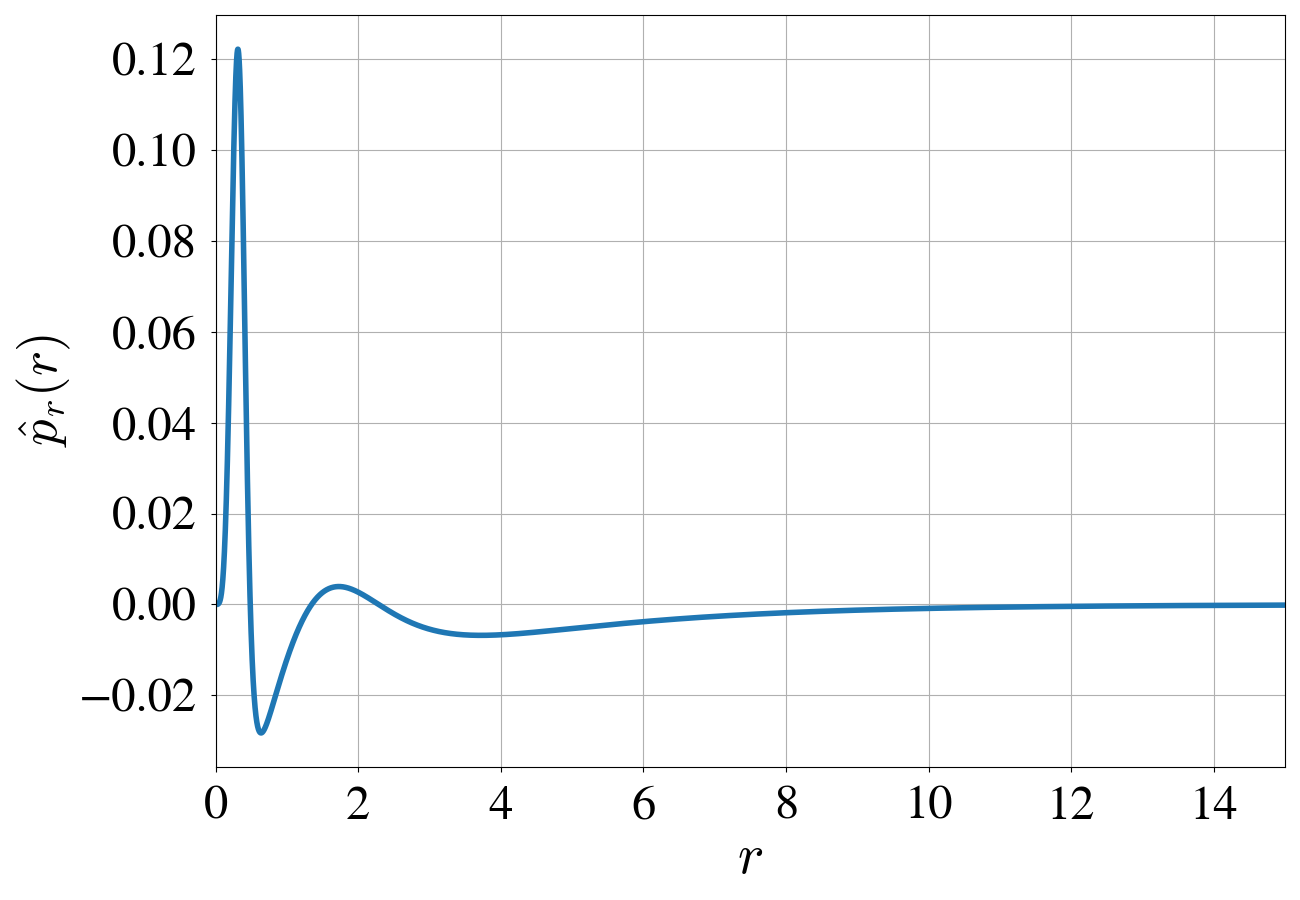}};
    \node[eqn] at ([yshift=-2mm]frame2.north) {$\kappa = 4,\,\,z=27$};
\end{tikzpicture}
    \caption{Two solutions with four fermions, $\kappa=4$, which both have distinct regions with negative radial pressure.}\label{fig:p_neg}
\end{figure}
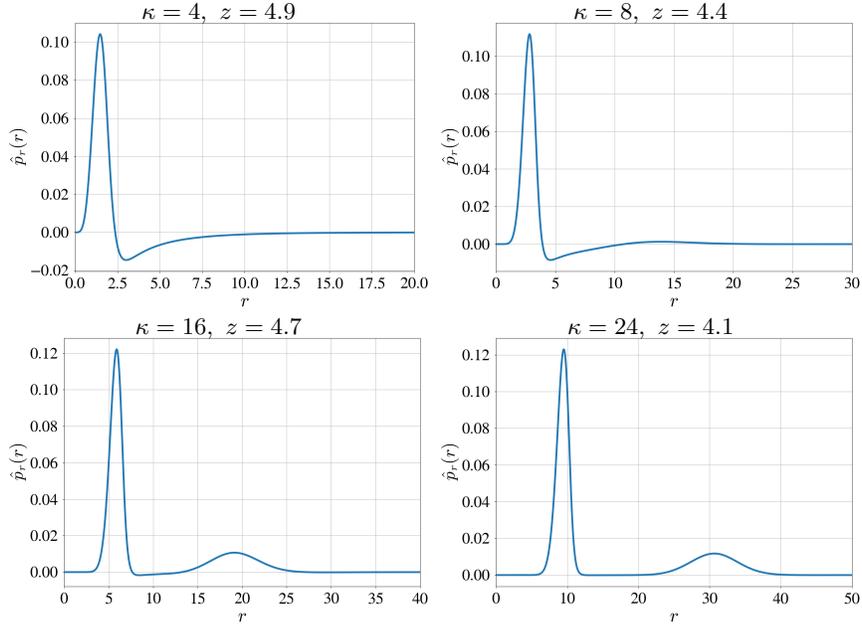
\begin{figure}[htb]
    \centering
    \begin{tikzpicture}[
        image/.style = {text width=0.45\textwidth, 
                        inner sep=0pt, outer sep=0pt},
        eqn/.style = {anchor=south, text=black, font=\footnotesize},
        node distance = 2.5mm and 0.5mm
    ]  
\node [image] (frame1)
    {\includegraphics[width=\linewidth]{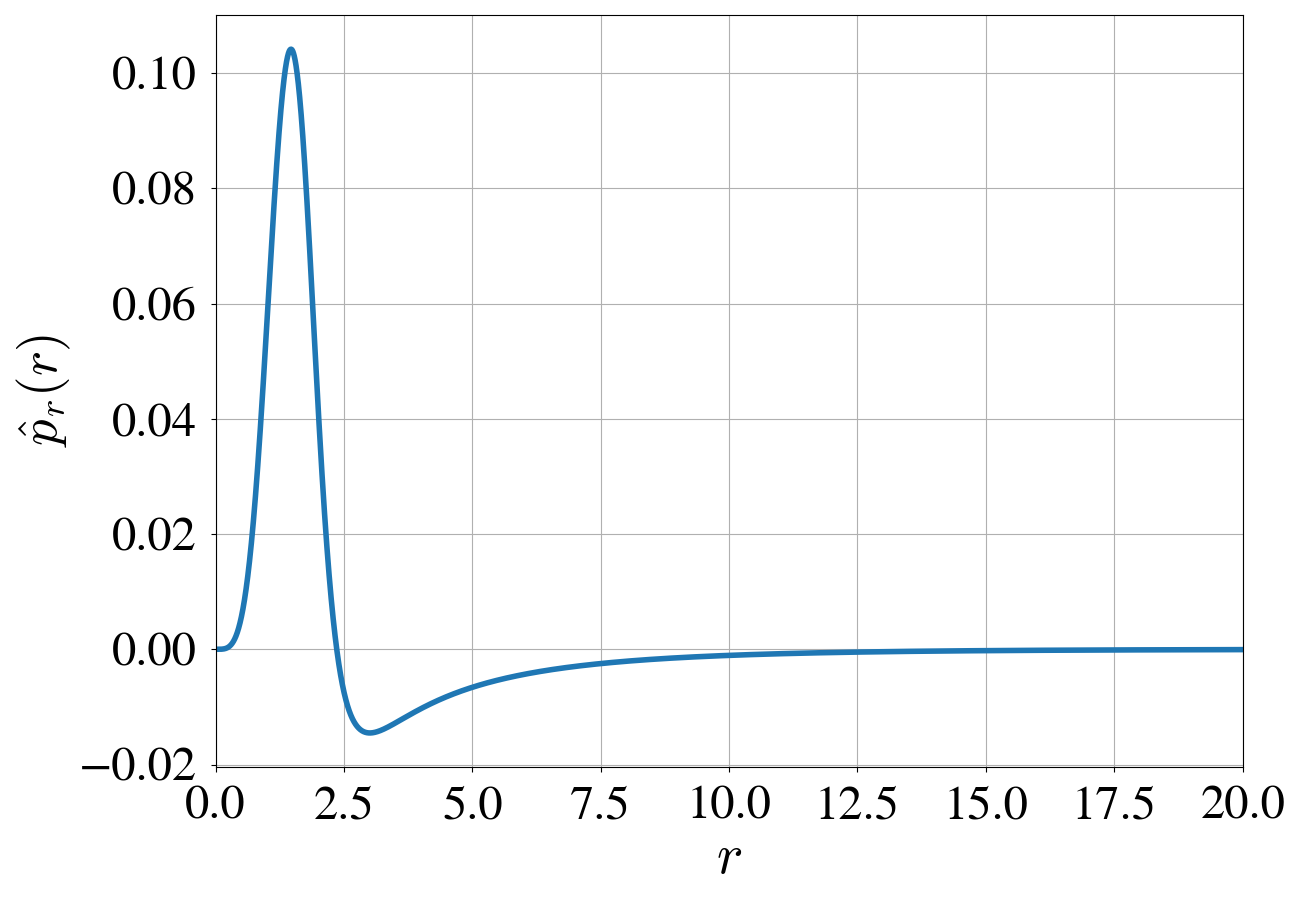}};
    \node[eqn] at ([yshift=-2mm]frame1.north) {$\kappa = 4,\,\,z=4.9$};
\node [image,right=of frame1] (frame2) 
    {\includegraphics[width=\linewidth]{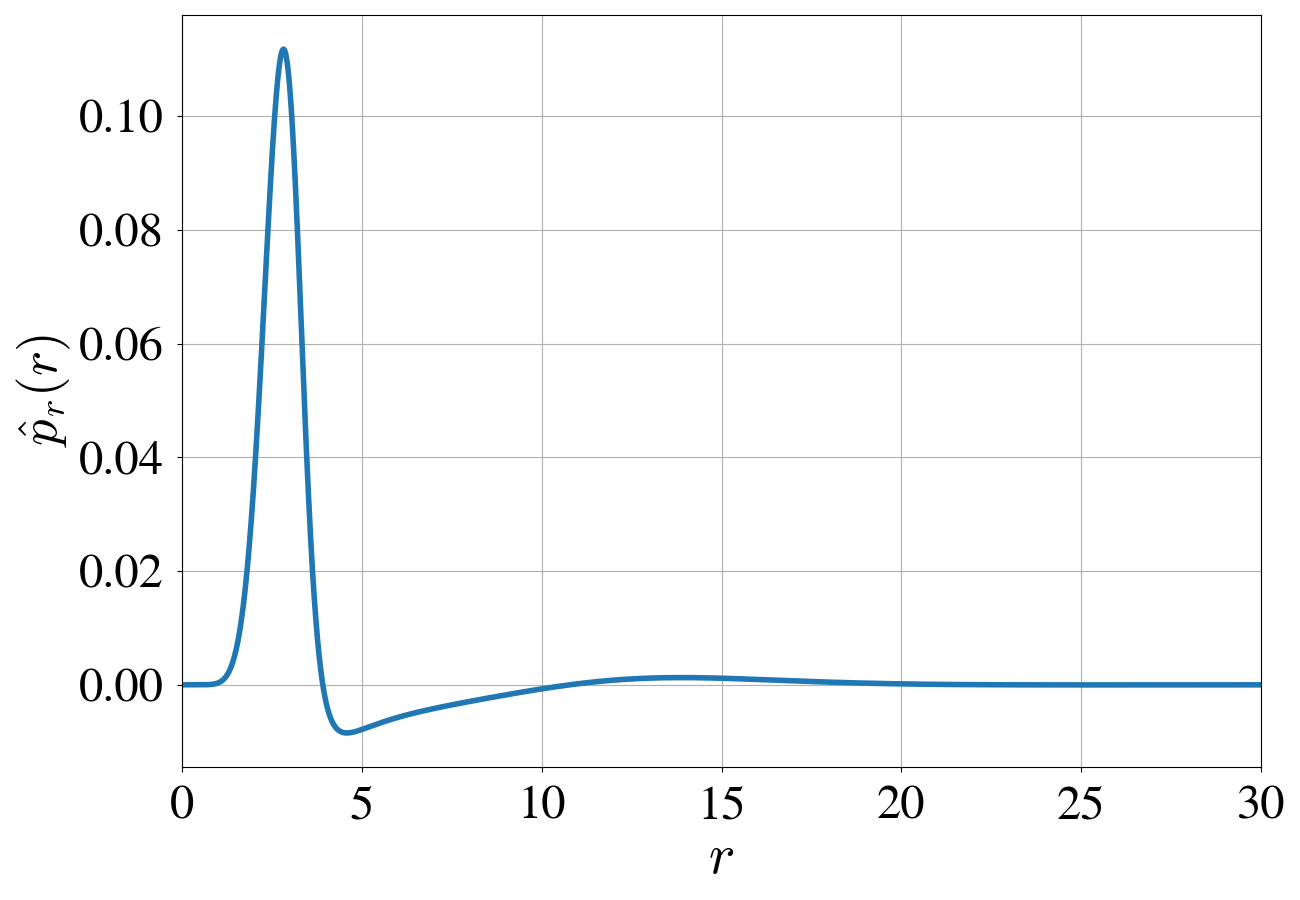}};
    \node[eqn] at ([yshift=-2mm]frame2.north) {$\kappa = 8,\,\,z=4.4$};
\node[image,below=of frame1] (frame3)
    {\includegraphics[width=\linewidth]{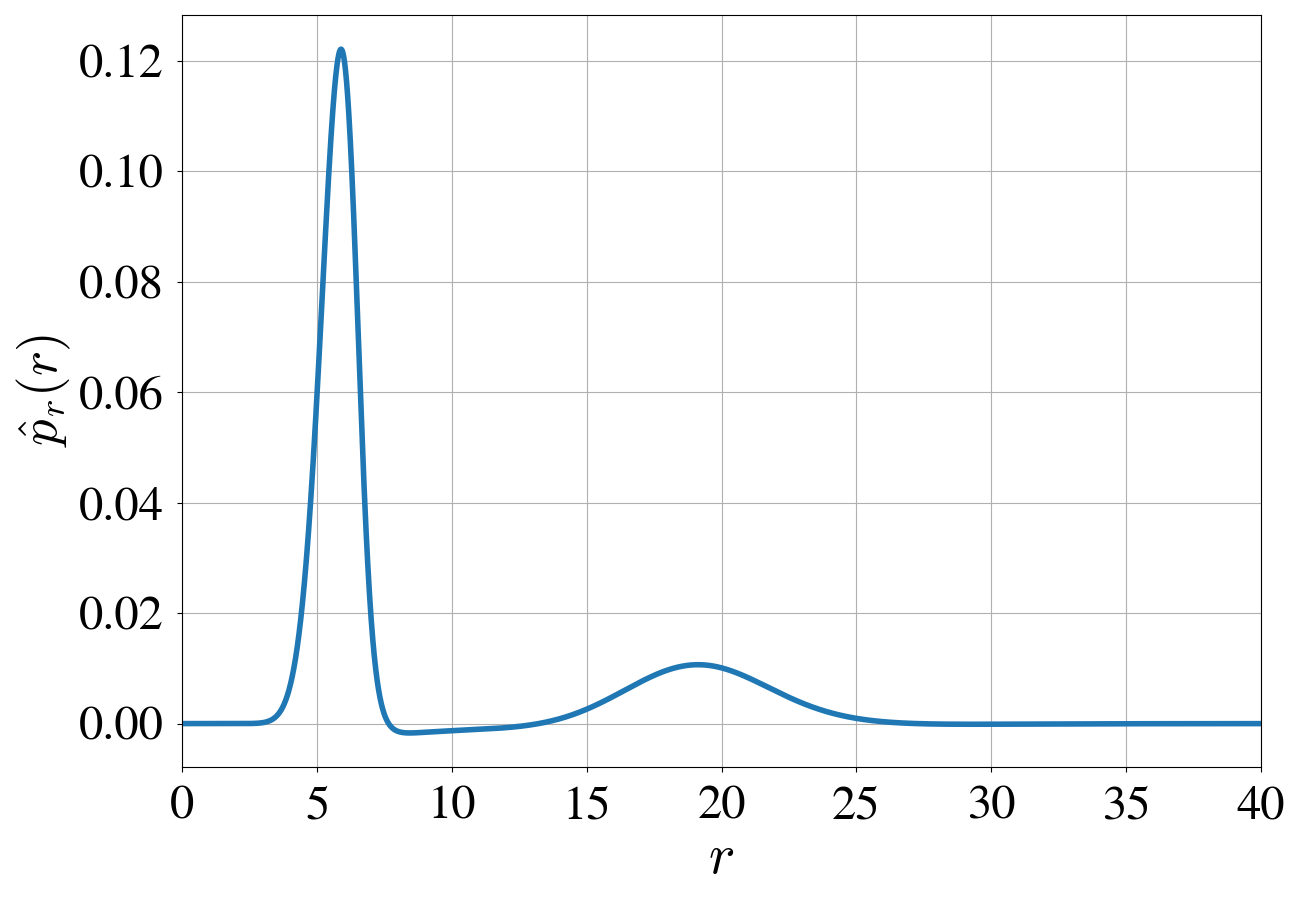}};
    \node[eqn] at ([yshift=-2mm]frame3.north) {$\kappa = 16,\,\,z=4.7$};
\node[image,right=of frame3] (frame4)
    {\includegraphics[width=\linewidth]{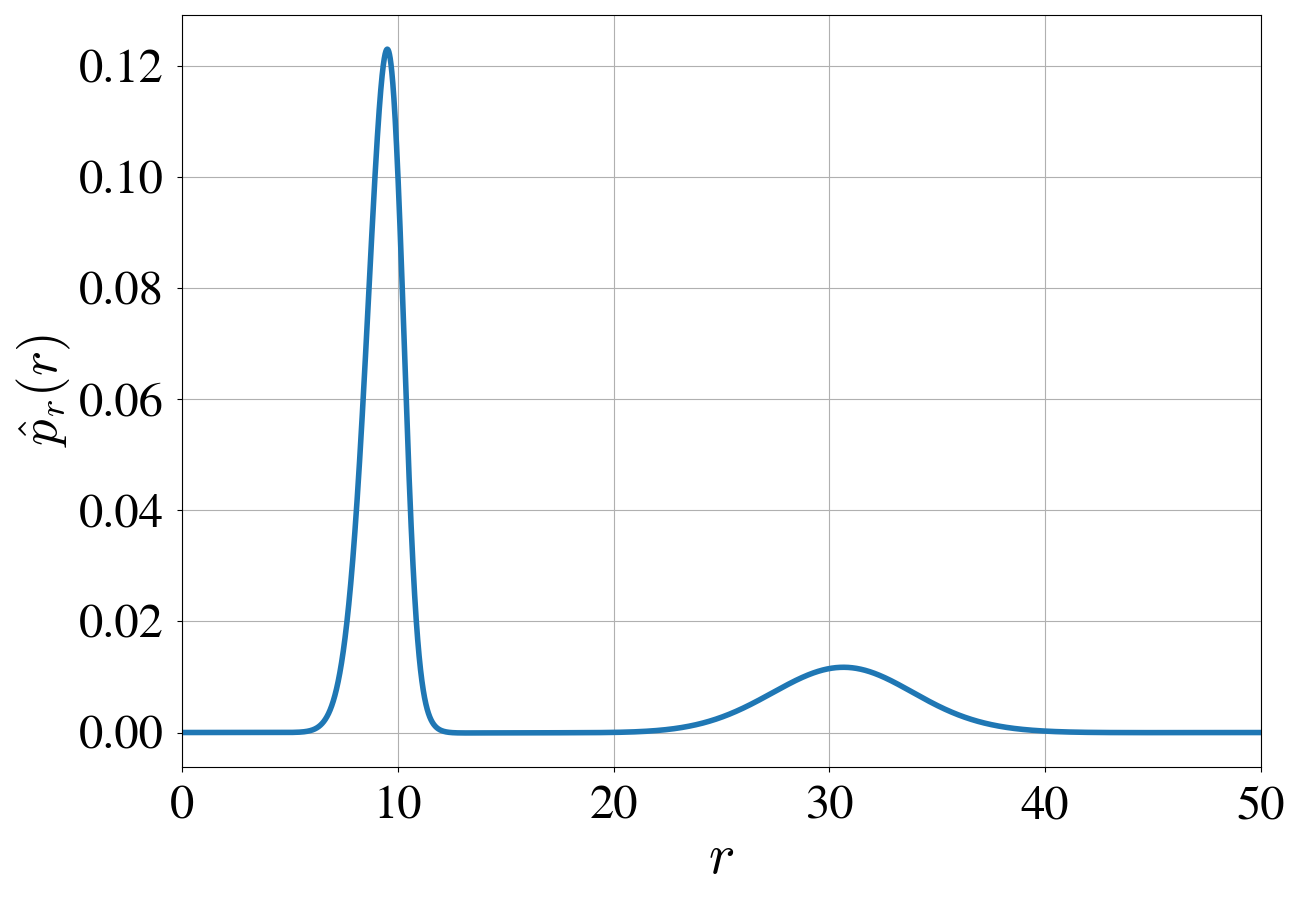}};
    \node[eqn] at ([yshift=-2mm]frame4.north) {$\kappa = 24,\,\,z=4.1$};
\end{tikzpicture}
    \caption{The radial pressure is shown for four solutions corresponding to $\kappa=4,8,16,24$. In all cases $z$ is between 4 and 5. The radial pressure becomes non-negative when the number of fermions increases ($16\leq \kappa \leq 24$) and we interpret this as a transition to phenomenological behaviour.}\label{fig:p_neg_seq1}
\end{figure}
\begin{figure}[htb]
    \centering
    \begin{tikzpicture}[
        image/.style = {text width=0.45\textwidth, 
                        inner sep=0pt, outer sep=0pt},
        eqn/.style = {anchor=south, text=black, font=\footnotesize},
        node distance = 2.5mm and 0.5mm
    ] 
\node [image] (frame1)
    {\includegraphics[width=\linewidth]{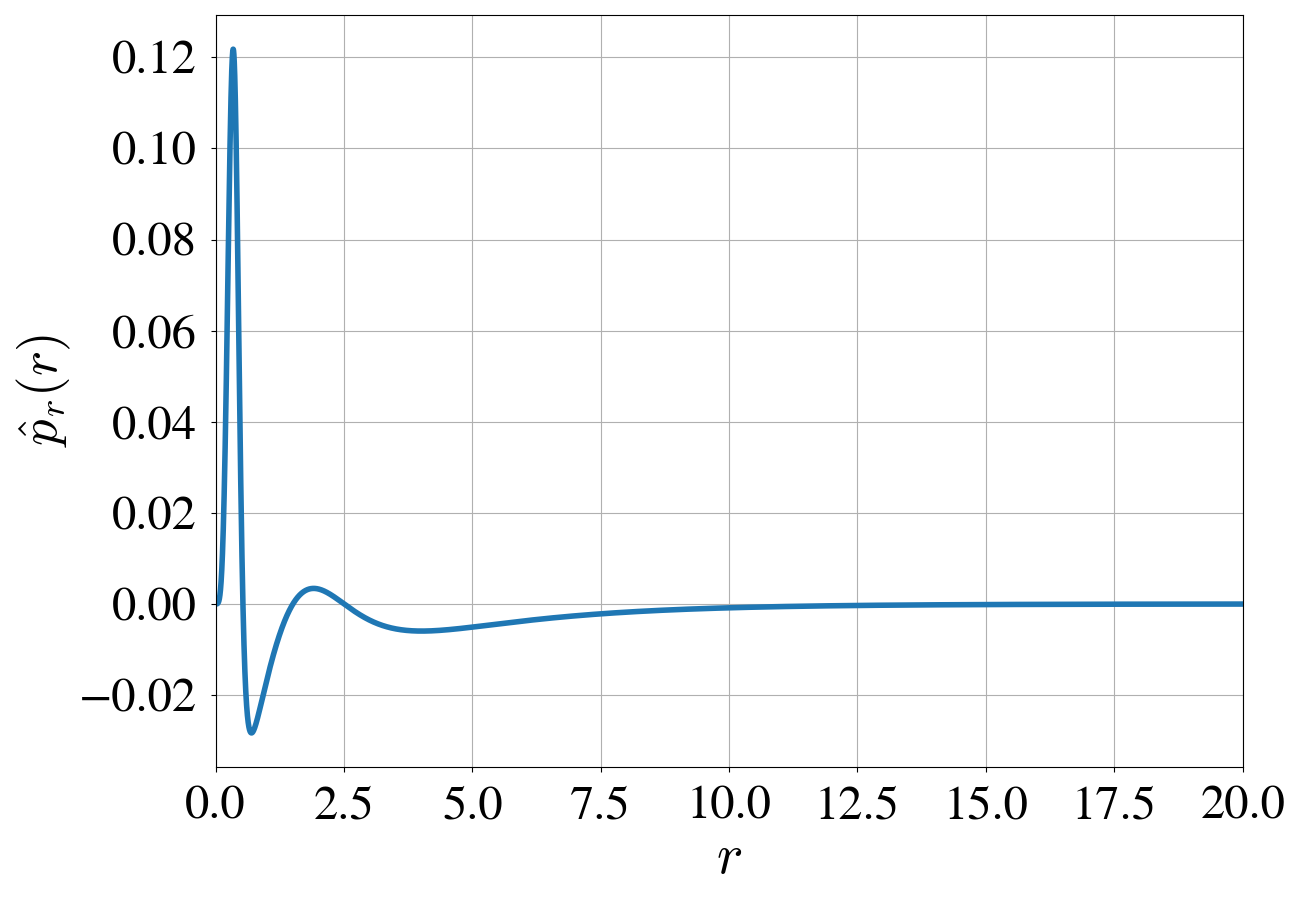}};
    \node[eqn] at ([yshift=-2mm]frame1.north) {$\kappa = 4,\,\,z=25$};
\node [image,right=of frame1] (frame2) 
    {\includegraphics[width=\linewidth]{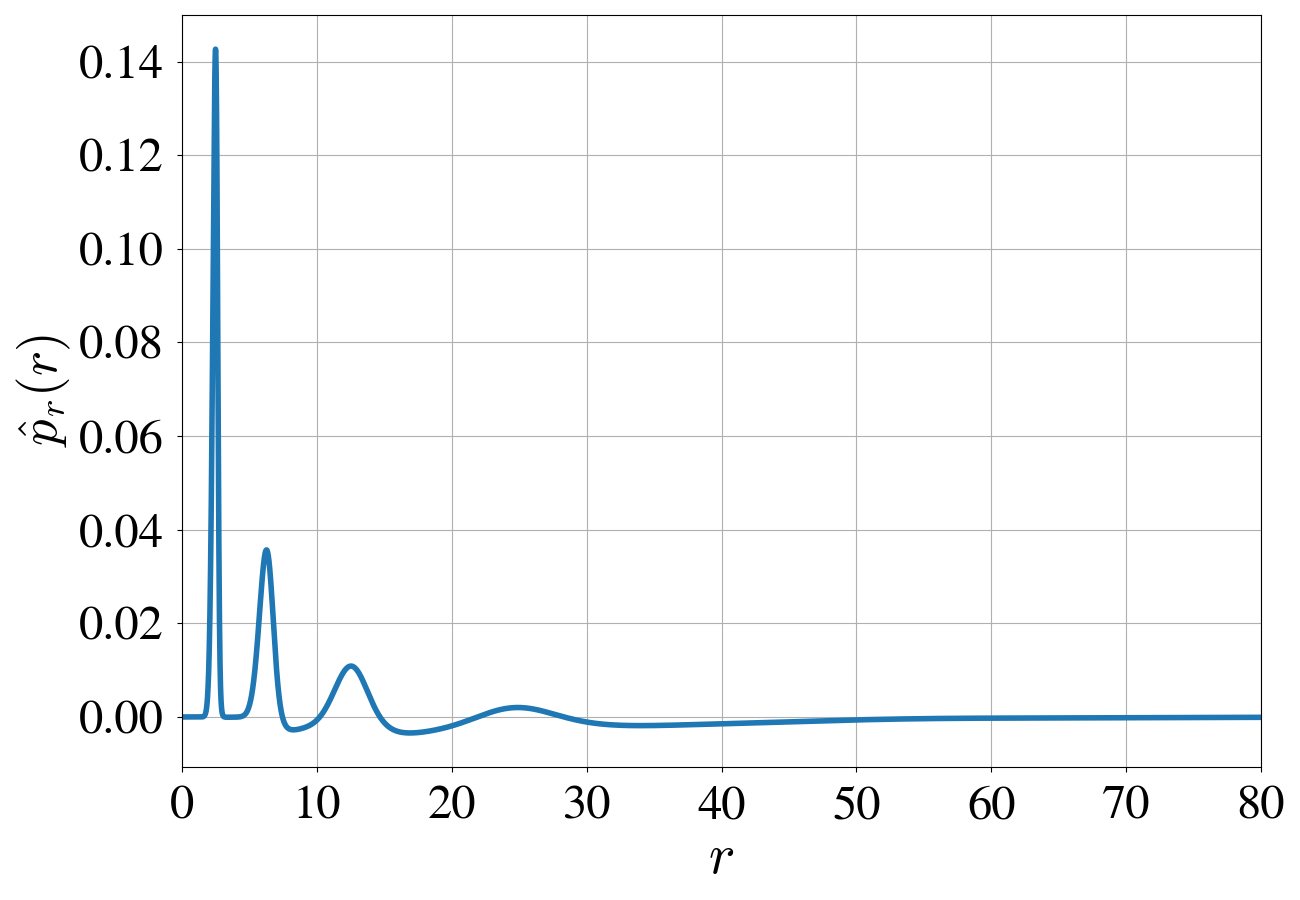}};
    \node[eqn] at ([yshift=-2mm]frame2.north) {$\kappa = 32,\,\,z=24$};
\node[image,below=of frame1] (frame3)
    {\includegraphics[width=\linewidth]{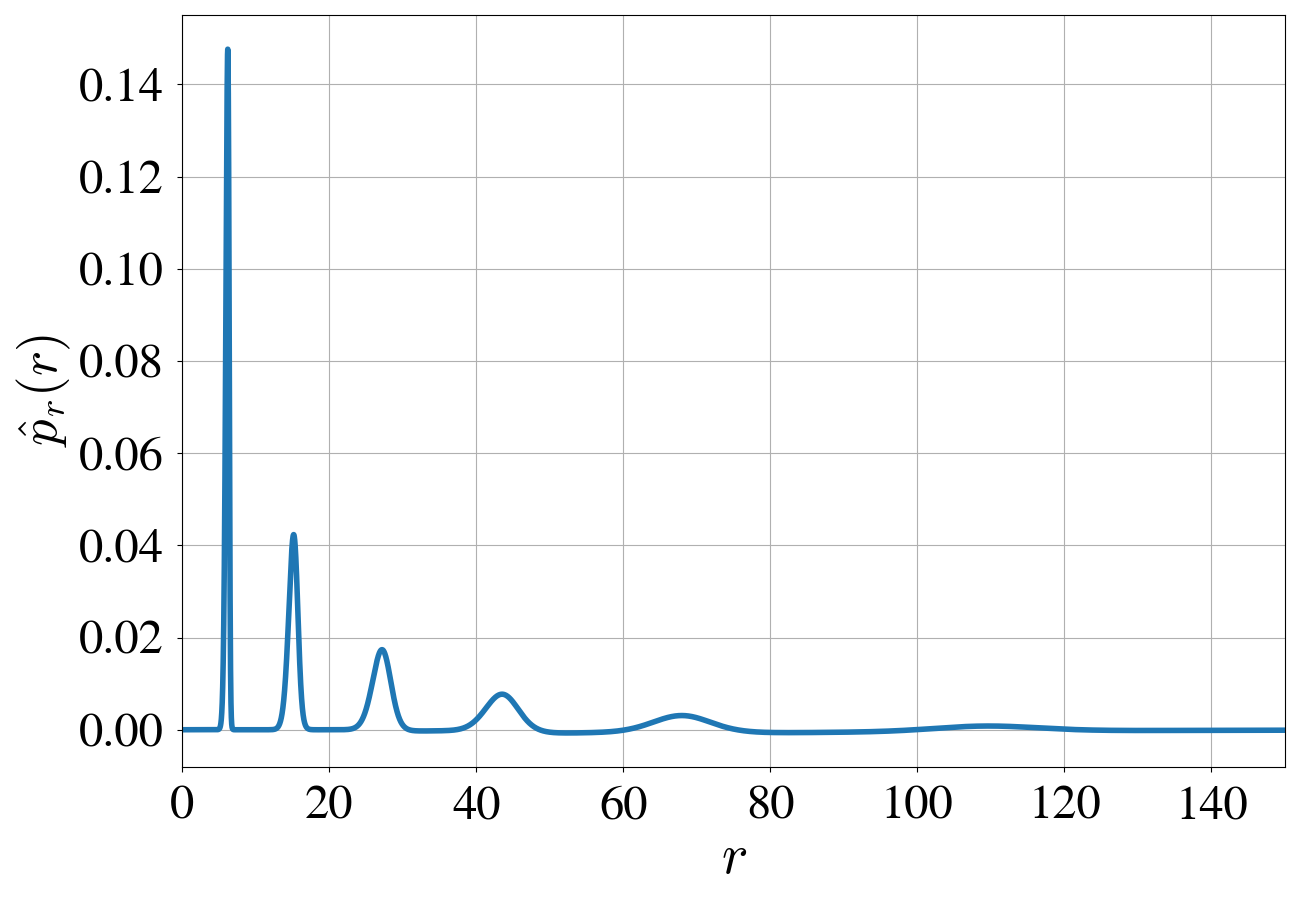}};
    \node[eqn] at ([yshift=-2mm]frame3.north) {$\kappa = 90,\,\,z=25$};
\node[image,right=of frame3] (frame4)
    {\includegraphics[width=\linewidth]{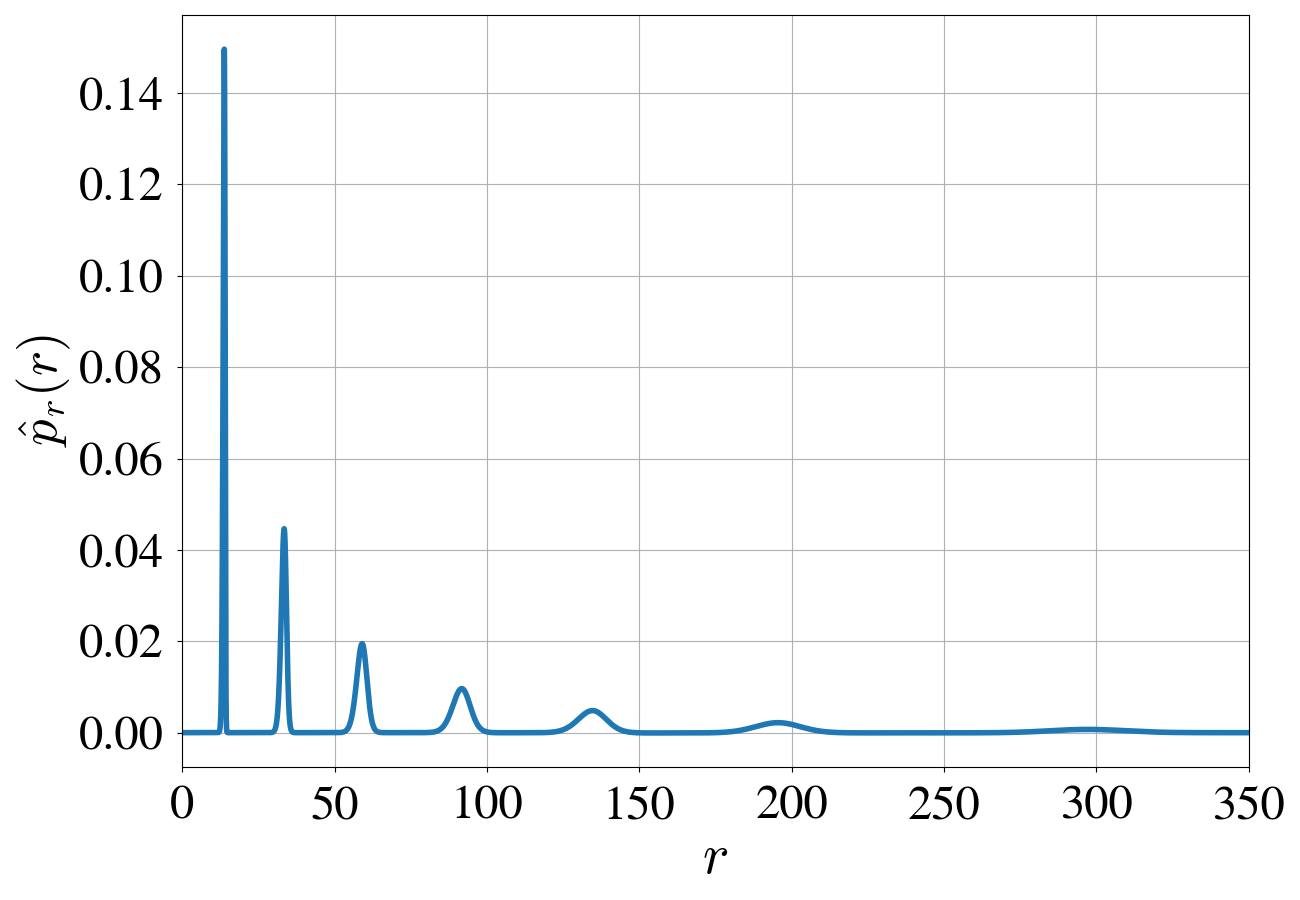}};
    \node[eqn] at ([yshift=-2mm]frame4.north) {$\kappa = 180,\,\,z=24$};
\end{tikzpicture}
    \caption{The radial pressure is shown for solutions corresponding to $\kappa=4,32,90,180$ with central redshift $z\approx 25$. As $z$ increases, a larger value of $\kappa$ is required for the radial pressure to become non-negative.}\label{fig:p_neg_seq2}
\end{figure}

For Vlasov matter and other phenomenological matter models the pressure is always non-negative everywhere, whereas field theoretical matter models may have regions where the pressure is negative. We therefore find it natural to investigate whether there are regions with negative pressure for solutions of the Einstein-Dirac system. The pressure is anisotropic for the solutions we construct and we distinguish between the radial pressure $p_r$ and the tangential pressure $p_{\perp}$. For all solutions that we have constructed, $p_{\perp}\geq 0$, but we do find solutions possessing regions for which $p_r<0$ so we focus here on the radial pressure. In Figure \ref{fig:p_neg} two solutions are shown for which there are distinct regions where $p_r<0$. In both cases $\kappa=4$, while the central redshift is $z=4.9$ and $z=27$, respectively. In the former case, the negative pressure region is one connected interval, while in the latter case it consists of two such intervals. The modulus of the pressure in these regions is comparatively small, but nevertheless it is negative, a property that we associate with the semiclassical nature of the Einstein-Dirac system. \textit{A central question that we pose in this work is how this property changes as the number of fermions increases.} Note that the number of fermions in Figure \ref{fig:p_neg} is small and we ask what happens to the sign of $p_r$ for a sequence of solutions with a similar central redshift but with increasing $\kappa$. In Figure \ref{fig:p_neg_seq1}, a solution sequence for which $z$ is approximately 4 is shown. The negative pressure region \textit{disappears} as $\kappa$ increases, that is, when $\kappa$ is between 16 and 24. In Figure \ref{fig:p_neg_seq2} 
the same phenomenon is also observed when the central redshift is higher ($z\approx 25$). In this case $\kappa$ has to be taken larger for the negative regions to disappear. We have considered sequences of solutions with even higher central redshift, and the same conclusion holds, that is, as $z$ increases, $\kappa$ must be taken larger for the negative regions to disappear. 

Since the Einstein-Dirac system is semiclassical, it can be said to have a quantum signature, and we think of the negative pressure regions as an effect originating from the quantum nature of Dirac's equation. As the number of fermions increases, this effect ceases to exist, and we interpret this as a transition to classical, phenomenological behaviour. We find it remarkable that the transition occurs for a comparatively low number of fermions.

Let us also include a discussion about another field-theoretical matter model. 
As we mentioned in the introduction, a similar behaviour of the pressure as for the Einstein-Dirac system occurs in the case of $\ell$-boson stars. 
Negative pressure regions exist in this case, see \cite[Fig. 5]{AS1}, and the modulus of the pressure in these regions is relatively small. This indicates that the pressure may generically behave in this way for a field-theoretical Einstein-matter system. If so, it would cast strong doubt on the choice $p=-\rho$ which is often found in the literature. 

We finish this section with a remark on the semiclassical interpretation of the Einstein-Dirac system. In the work \cite{Kn} the relation between a quantized Dirac field and its semiclassical analogue is analyzed. This work was inspired by a similar study \cite{AS2} on boson stars.

\section{Highly compact solutions and bounds on $2m/r$}\label{sec-Gamma}
An interesting question is what the value of the maximum compactness $\Gamma$ is for solutions of the Einstein-Dirac system, and how $\Gamma$ depends on the number of particles $\kappa$. Naively, one could invoke the well-known Buchdahl bound \cite{Bu} which states that a static solution for which the energy density is monotonically decreasing outwards, and for which the pressure is isotropic, has the property that
\begin{equation}\label{Buchdahl} 
\frac{2M}{R}\leq 8/9,
\end{equation} 
where $M$ is the total mass and $R$ is the radius of the boundary of the object. (The ratios $2M/R$ and $\Gamma$ are slightly different quantities, but clearly $2M/R\leq\Gamma$.) However, the solutions we study in this work do not share the properties of an isotropic fluid such that the energy density decreases monotonically outward and the pressure is isotropic. In contrast, the multi-peak structure of the energy density is as far from monotonic as one can imagine. Moreover, the pressure is highly anisotropic, with the tangential pressure dominating the radial pressure for compact solutions; cf. Figure \ref{fig:all_T}. Hence, the assumptions used by Buchdahl to derive his bound are manifestly violated for the solutions that we consider in this work. 
In \cite{A2} it was shown by the first author that any static spherically symmetric solution of the Einstein-matter system which satisfies 
\begin{equation}\label{energyconditionOmega}
p_r+2p_t\leq \Omega\rho, 
\end{equation}
where $\Omega$ is a positive constant,
and for which the radial pressure $p_r$ is non-negative, satisfies the bound
\begin{equation}\label{Abound1}
\Gamma=\sup\frac{2m(r)}{r}\leq\frac{(1+2\Omega)^2-1}{(1+2\Omega)^2}.
\end{equation}
The bound is sharp in the following sense. Consider a sequence of shell solutions, indexed by $n$, with spatial support on the interval $[R_0^n,R_1^n]$, where 
\begin{equation}\label{R01ratio}
\frac{R_1^n}{R_0^n}\to 1 \,\, \textrm{as }\,\, n\to \infty.
\end{equation}
Then we have \cite{A2}
\[
\lim_{n\to\infty}\Gamma_n=\frac{(1+2\Omega)^2-1}{(1+2\Omega)^2},
\]
where $\Gamma_n$ denotes the maximum compactness of the solution corresponding to the index $n$. 
An infinitely thin shell solution could thus be said to saturate the bound. However, an infinitely thin shell is not a regular solution, and we therefore prefer to think of sharpness in the sense just described, i.e., a sequence of regular solutions which approaches the upper bound. With abuse of notation, we still write that an infinitely thin shell solution satisfies the bound.

If $\Omega=1$, then the energy condition (\ref{energyconditionOmega}) reads 
\begin{equation}\label{energycondition}
p_r+2p_t\leq \rho,
\end{equation}
and the bound (\ref{Abound1}) becomes 
\begin{equation}\label{Abound2}
\Gamma\leq 8/9.
\end{equation} 
For solutions of the Einstein\,-Vlasov system it always holds that $\Omega=1$, and that $p_r\geq 0$, so both (\ref{energycondition}) and (\ref{Abound2}) are valid in this case. Note that (\ref{Abound2}) coincidently gives the same value $\frac89$ as in the Buchdahl bound (\ref{Buchdahl}). However, even if the upper bounds are the same, the inequalities are very different as mentioned above. Furthermore, the saturating solution of (\ref{Buchdahl}) is the Schwarzschild interior solution with constant energy density, for which the pressure diverges in the centre. This solution thus violates the dominant energy condition and is not a physically realistic solution. For inequality (\ref{Abound1}) the saturating solution is an infinitely thin shell (in the sense described above). It satisfies the energy condition $p_r+2p_t\leq \Omega\rho$ and is sound from a physics point of view. In the work \cite{A2}, any matter model satisfying (\ref{energyconditionOmega}) was considered. Sharpness was proved under the \textit{hypothesis} that a sequence of shell solutions exist with the property (\ref{R01ratio}). It is not clear that such a sequence in fact exists for a coupled Einstein-matter system. In the case of the Einstein,-Vlasov system, the first author \cite{A1} proved that shell solutions indeed exist that satisfy (\ref{R01ratio}). Hence, the inequality (\ref{Abound2}) is sharp for solutions to the Einstein\,-Vlasov system. 

Some natural questions arise about solutions to the Einstein-Dirac system. The first question is whether or not the solutions satisfy the energy condition (\ref{energyconditionOmega}) for some $\Omega$, and in particular in the case $\Omega=1$? In order to use the bound (\ref{Abound1}), not only (\ref{energyconditionOmega}) must hold, but also the radial pressure must be non-negative. We found in Section \ref{sec-pressure} that the pressure is non-negative when $\kappa$ is sufficiently large. Hence, we are mainly interested in checking the condition (\ref{energyconditionOmega}) when $\kappa$ is comparatively large. We emphasize that $\Gamma$ increases as $\kappa$ and $z$ increase, cf. Figure \ref{fig:Gamma}, so it is not a restriction to assume that $\kappa$ and $z$ are large to investigate $\Gamma$. 

\begin{figure}[htbp]
\begin{center}
\scalebox{.25}{\includegraphics{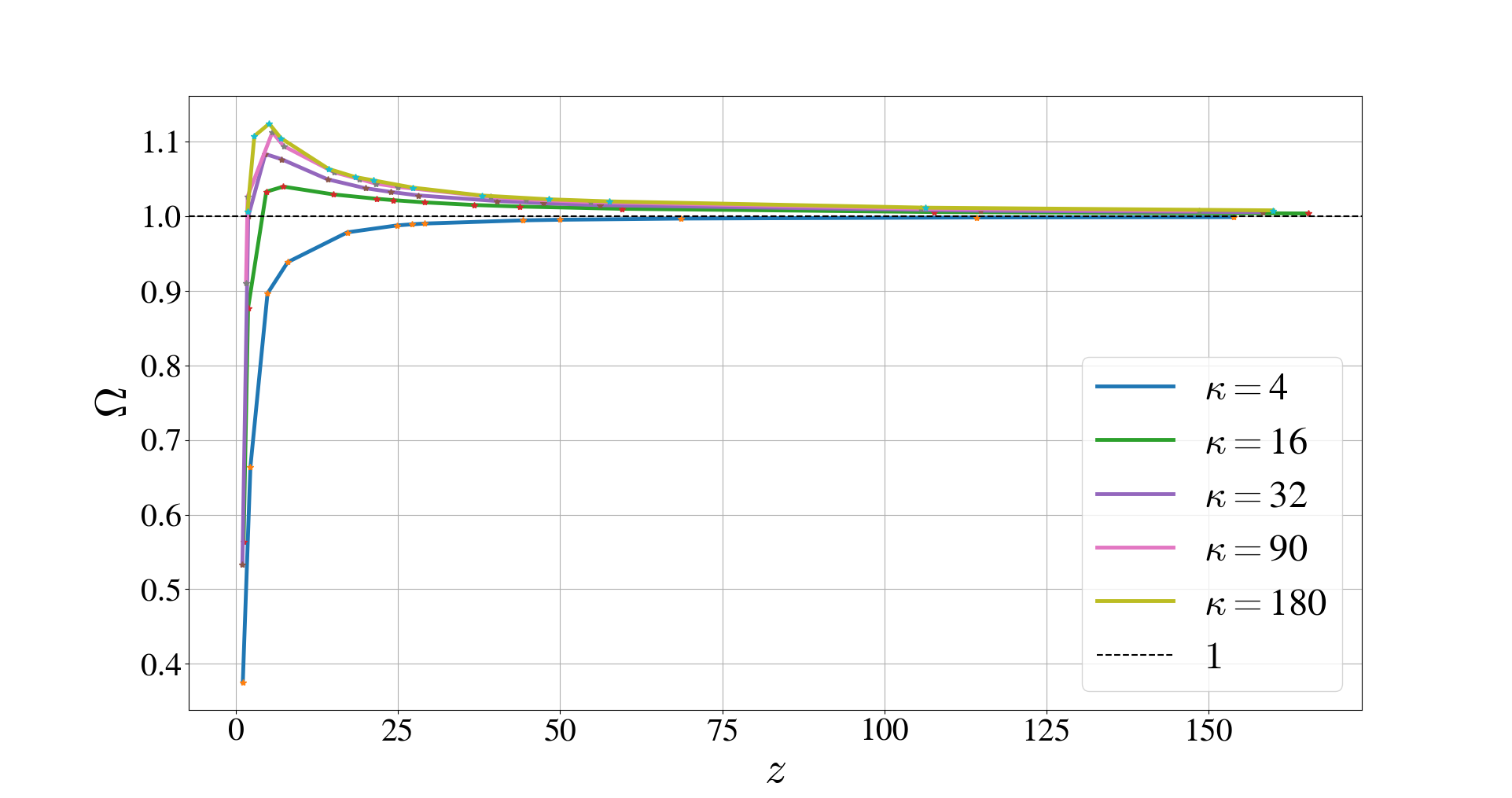}}
\end{center}
\caption{The maximum value of $\Omega(r)$ on the entire solution interval for different values of  $\kappa$ and $z$. Clearly, (\ref{energyconditionOmega}) holds with $\Omega=1.12$.}\label{fig:quotient-all}
\end{figure}

\begin{figure}[htbp]
\begin{center}
\scalebox{.25}{\includegraphics{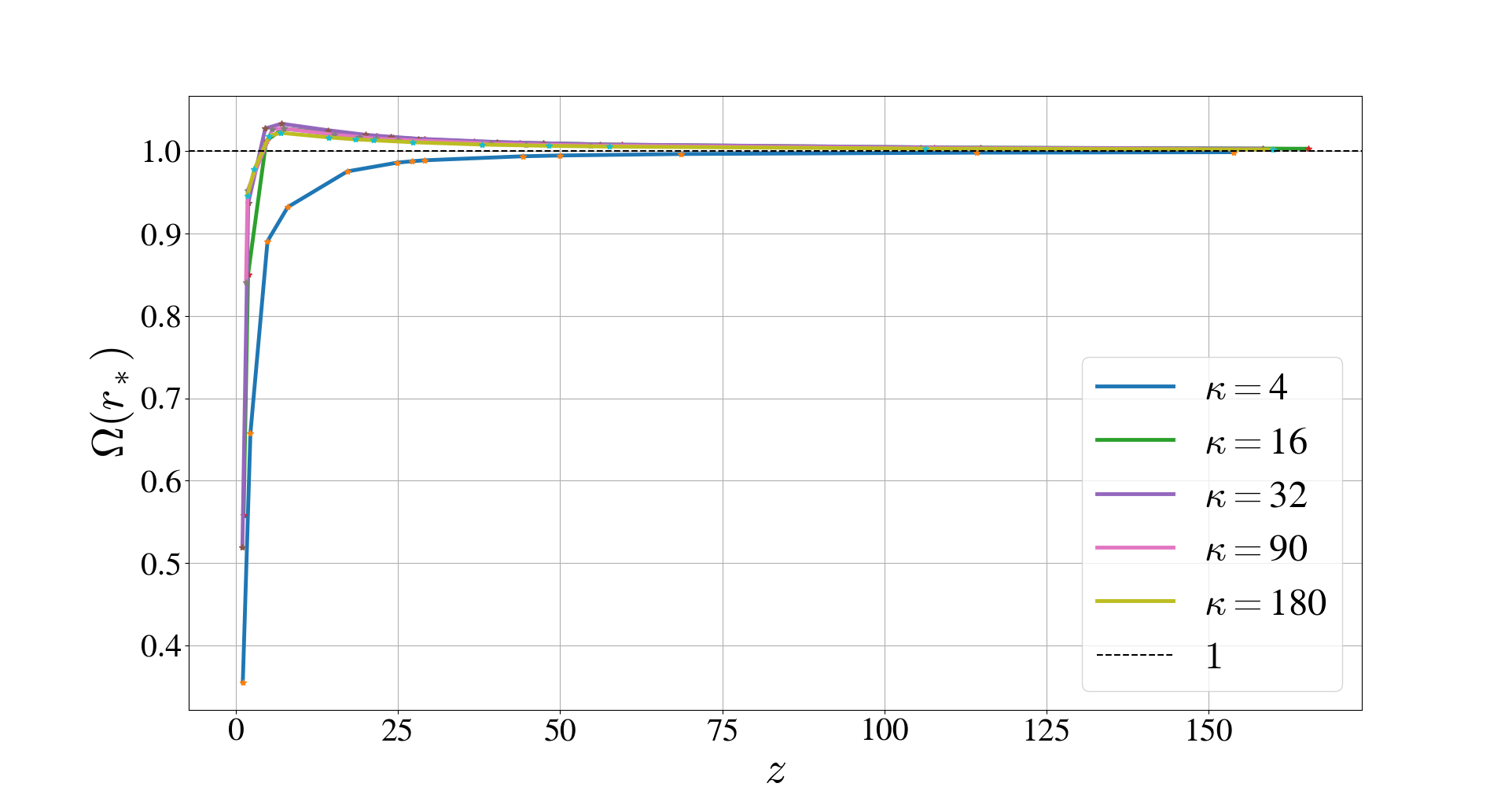}}
\end{center}
\caption{The value of $\Omega(r_*)$ for different values of  $\kappa$ and $z$. We find that $\Omega(r_*)\leq 1.02$ and as $z$ increases $\Omega(r_*)\to 1$.}\label{fig:quotient-rstar}
\end{figure}

\begin{figure}[htbp]
\begin{center}
\scalebox{.25}{\includegraphics{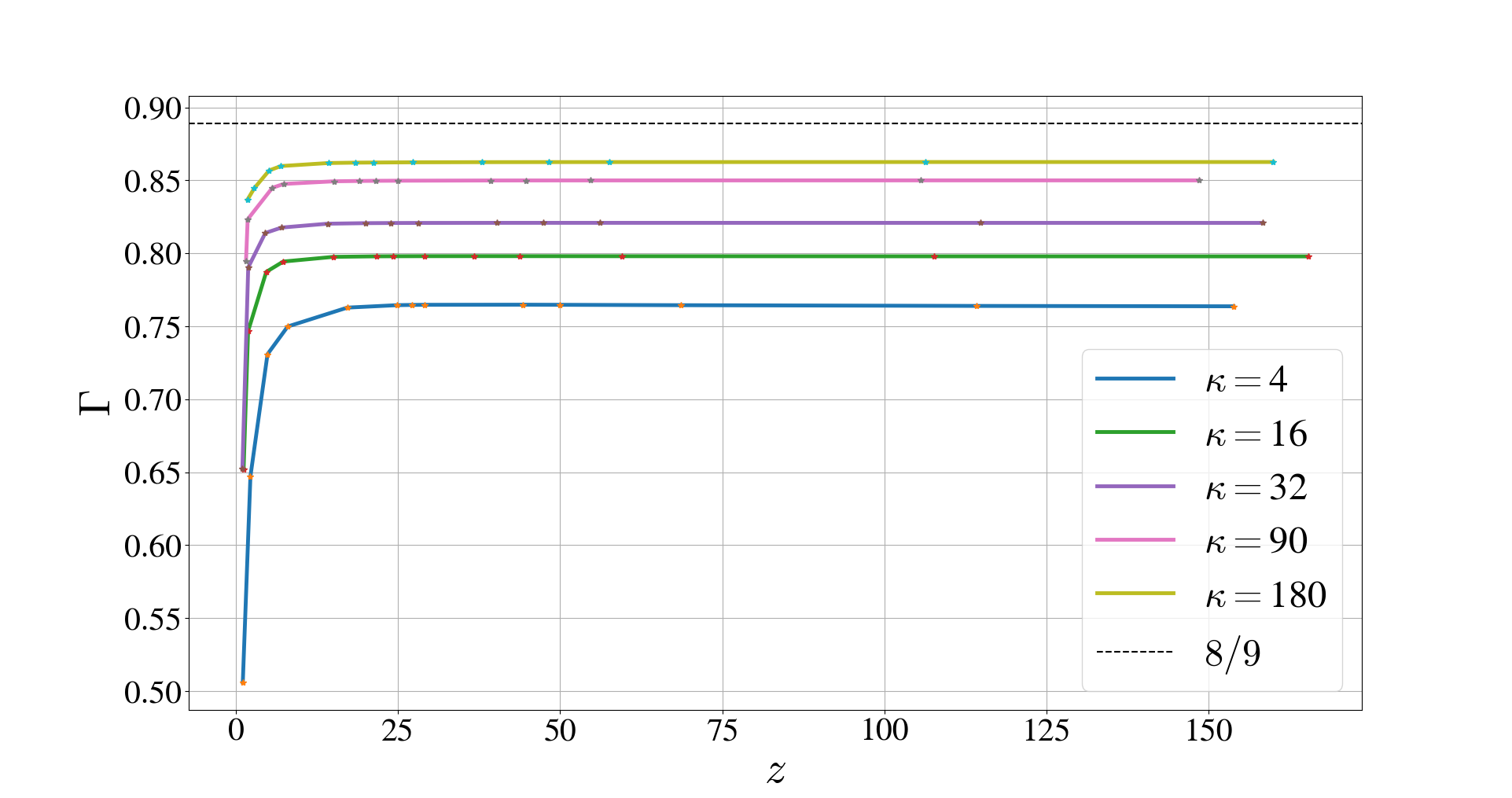}}
\end{center}
\caption{The value of $\Gamma$ for different values of $\kappa$ and $z$. $\Gamma$ increases as $\kappa$ increases and the question is if $\Gamma\to\frac89$ as $\kappa\to\infty$.}\label{fig:Gamma}
\end{figure}

%
%
%
We define
\begin{equation}\label{Omegar}
\Omega(r):=\frac{p_r(r)+2p_t(r)}{\rho(r)},
\end{equation}
so that $\Omega=\sup_r\,\Omega(r)$. For the solutions that we consider in this work, the maximum compactness is attained at a finite radius, which we denote by
$r_*$, that is, 
\[ 
\Gamma =\frac{2m(r_*)}{r_*}.
\]
It is located in the peak closest to the centre. It is natural to check the condition (\ref{energyconditionOmega}) at $r=r_*$ and at any value of $r$ in the solution interval. In Figure \ref{fig:quotient-all}, $\Omega$ is shown, that is, the maximum value of $\Omega(r)$ throughout the solution interval. We find that (\ref{energyconditionOmega}) holds for $\Omega\leq 1.12$. 
Note that as the central redshift $z$ increases, $\Omega$ approaches 1 independently of $\kappa$. The maximum value in Figure \ref{fig:quotient-all} is in fact attained far out where the energy density of the solution is very small. Hence, we find that the ratio at $r=r_*$ is the relevant case for studying the maximum compactness ratio $\Gamma$. In Figure \ref{fig:quotient-rstar} it is found that (\ref{energyconditionOmega}) holds at $r=r_*$ for $\Omega\leq 1.02$, and as $z$ increases, $\Omega$ approaches 1 independently of $\kappa$. In Figure \ref{fig:Gamma} $\Gamma$ is depicted for the solutions studied in Figures \ref{fig:quotient-all} and \ref{fig:quotient-rstar}. 
As $z$ increases, $\Gamma$ seems to be fixed at a constant value, but by checking the numerical values, it follows that $\Gamma$ increases with $z$.

Consequently, we conclude that (\ref{energyconditionOmega}) is consistent with $\Omega=1$ in the compact regime, that is, in the regime where $2m(r)/r$ is large. The highly compact solutions of the Einstein-Dirac system thus obey the same energy condition (\ref{energyconditionOmega}) as the solutions of the Einstein,-Vlasov system. Since the radial pressure for these solutions is non-negative, \textit{both hypotheses required for the validity of the bound (\ref{Abound2}) hold also for compact solutions of the Einstein-Dirac system}. If compact solutions of the Einstein-Dirac system and of the Einstein\,-Vlasov system should share exactly the same properties regarding their compactness, then it is necessary that (\ref{Abound2}) is sharp for solutions of both systems. Hence, the question is whether or not a sequence of solutions to the Einstein-Dirac system exists such that in the limit $\Gamma=\frac89$. This question is investigated in the next section.

Let us end the discussion on the validity of (\ref{Abound2}) by asking a similar question in the charged case. From a physics perspective it would be reasonable to consider charged fermions rather than neutral fermions. Finster, Smoller and Yau extended their study \cite{FSY1} to the case of charged fermions in \cite{FSY2}. They found static solutions of the Einstein-Dirac-Maxwell system in the case of two particles. In the spirit of the work of Leith et al. \cite{LHHD} and the present work, it would be natural to generalize the study \cite{FSY2} to the case of many charged fermions. It would then be very interesting to investigate the maximum compactness of solutions to the Einstein-Dirac-Maxwell system. Indeed, the bound (\ref{Abound2}) was generalized to the charged case in \cite{A4}
\begin{equation}\label{charged_ineq}
\sqrt{m(r)}\leq\frac{\sqrt{r}}{3}+\sqrt{\frac{r}{9}+\frac{q^2(r)}{3}},
\end{equation}
where $q(r)$ and $m_g(r)$ is the charge and gravitational mass within area radius $r$, respectively. 
In the neutral case, there is a gap between $\frac89$ and 1, i.e. static solutions cannot be constructed which are arbitrarily "close" to Schwarzschild black hole solutions. In the charged case, there is no gap, and solutions exist that are arbitrarily "close" to a Reissner-Nordström solution. The validity of the bound (\ref{charged_ineq}) requires that the energy condition (\ref{energycondition}) is satisfied and that $p_r\geq 0$. It would be interesting to check whether this condition is satisfied for compact solutions of the Einstein-Dirac-Maxwell system. If so, it is motivated to explore if solutions can be constructed that saturate the bound (\ref{charged_ineq}), i.e. if static solutions of this system exist that are arbitrary close to an (extremal) Reissner-Nordstrom black hole solution. In the case of the Einstein-Vlasov-Maxwell system, solutions that saturate (\ref{charged_ineq}) were numerically constructed in \cite{AER}. Hence, there is good hope that similar solutions could also exist in the case of the Einstein-Dirac-Maxwell system with many particles.

%
%
%

\section{Sharpness of the bound $\Gamma\leq\frac89$}\label{sec-sharpness}
\begin{figure}[htbp]
\begin{center}
\scalebox{.25}
{\includegraphics{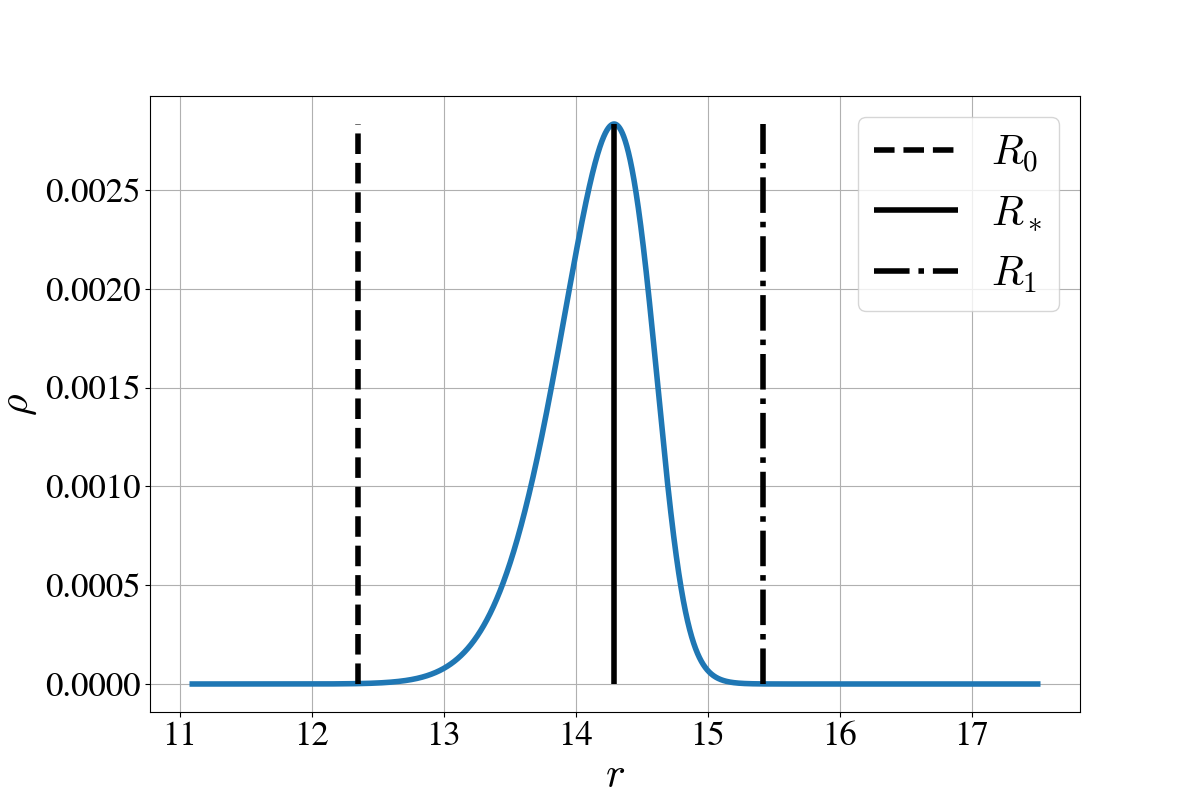}}
\end{center}
\caption{An example of the support $[R_0,R_1]$ computed by the algorithm.}\label{fig:ratio_algorithm}
\end{figure}

\begin{figure}[htbp]
\begin{center}
\scalebox{.25}{\includegraphics{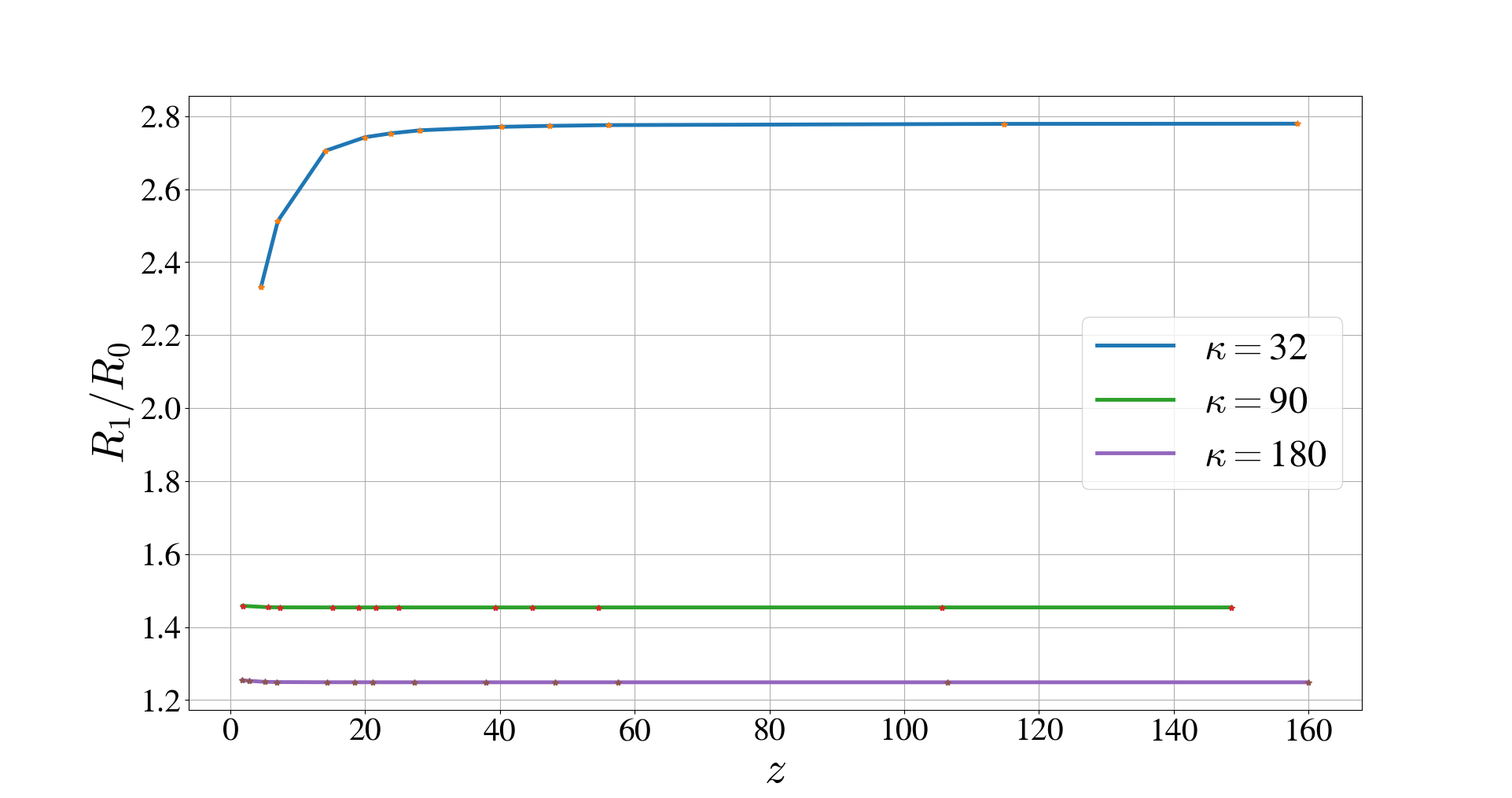}}
\end{center}
\caption{The ratio $R_1/R_0$ of the innermost peak of solutions to the Einstein-Dirac system for some different values of $\kappa$ and $z$. The question is if this ratio approaches 1 as $\kappa\to\infty$.}\label{fig:R1R0-ratio}
\end{figure}

It remains to determine how large $\Gamma$ \textit{actually} can be for solutions of the Einstein-Dirac system; the bound (\ref{Abound2}) is only an upper bound. As mentioned above, the bound is sharp if a sequence of shell solutions can be constructed such that the ratio $R_1^n/R_0^n$ tends to $1$ as $n\to\infty$, where $R_0^n$ and $R_1^n$ are the inner and outer radius of the shells in the sequence. For the Einstein\,-Vlasov system it is known \cite{A1} that such a sequence of shell solutions exists. In this case, the innermost shells of a compact multi-peak solution are separated by vacuum regions. Hence, it is well defined what we mean by the inner radius $R_0$ and the outer radius $R_1$, of the innermost shell for solutions of the Einstein\,-Vlasov system. 
\begin{remark}
The family of shells of the Einstein\,-Vlasov system constructed in \cite{A1} become more concentrated the closer to the centre they get, similar to the Einstein-Dirac solutions we consider here. In the limit, the inner radius $R_0\to 0$, whereas the outer radius $R_1$ is shown to satisfy $R_1\leq R_0+R_0^q$, where $q>1$. Hence $R_1/R_0\to 1$ as $R_0\to 0$. In the work \cite{A5} another class of shell solutions to the Einstein\,-Vlasov system is constructed, which surrounds a Schwarzschild black hole. In this case, the inner radius cannot approach $r=0$. Instead, $R_0\to \infty$ and $R_1$ satisfy $R_1\leq R_0+R_0^q$ where $q<1$. Hence $R_1/R_0\to 1$ as $R_0\to \infty$ and the maximum compactness ratio $\Gamma\to\frac89$ also in this situation. For the Einstein\,-Vlasov system, there are thus two different families of solutions, with the property that $\Gamma\to \frac89$. Whether a similar class of solutions with $R_0\to \infty$ exists for the Einstein-Dirac system, we are not aware. However, from the result \cite{FSY3} it is clear that static solutions of the Einstein-Dirac system that surround a black hole cannot exist.
\end{remark}
For multi-peak solutions of the Einstein-Dirac system, the peaks are not separated by complete vacuum regions; it is almost vacuum between the peaks. 
We have implemented the following strategy to determine $R_0$ and $R_1$ of the innermost shell in this case. First we evaluate the integral
\begin{equation}\label{extended_mass}
m_i:=4\pi \int_0^{R_2}s^2\rho(s)\, ds,
\end{equation}
where $R_2$ is chosen between the first and second peak where the energy density $\rho$ is very small. We then look for values $R_0>0$ and $R_1<R_2$, chosen about the radius where the maximum value of $\rho$ in the innermost shell is attained (which we denote by $R_*$), such that 
\begin{equation}\label{99mass}
4\pi \int_{R_0}^{R_1}s^2\rho(s)\, ds=0.9999 \, m_i.
\end{equation}
See Figure \ref{fig:ratio_algorithm} for an illustration of how $R_0$ and $R_1$ are chosen. (It is tacitly understood that we take the interval as small as possible.) 
In this way, we ensure that 99.99$\%$ of the mass of the innermost peak is enclosed between $R_0$ and $R_1$, and that $R_0<R_*<R_1$. Clearly, $R_0$ and $R_1$ are not uniquely determined, but the ratio $R_1/R_0$ is almost unchanged if the values of $R_0$ and $R_1$ vary slightly, while still preserving the enclosed mass between $R_0$ and $R_1$. It is common to use this type of criterion for field-theoretical solutions which do not vanish completely. In Figure \ref{fig:R1R0-ratio} the ratio $R_1/R_0$ is depicted for solutions corresponding to $\kappa=32,90,180$ for different values of $z$.

It is clear that the ratio decreases as $\kappa$ increases and reaches $1.2$ when $\kappa=180$. From the trend, it seems likely that the ratio decreases further as $\kappa$ increases, but it is numerically very challenging to increase $\kappa$ above 180. Whether $R_1^{\kappa}/R_0^{\kappa}\to 1$, as $\kappa\to \infty$, is therefore an open question. If the ratio tends to 1 then \cite{A2} guarantees that the inequality (\ref{Abound2}) is sharp. Assuming that it does not tend to 1; then it either tends to a value strictly larger than 1 or it does not tend to any specific value at all. In neither case are we able to determine the value $\Gamma$ tends to, that is, even if the ratio tends to a specific value (larger than 1), it is not straightforward to use the arguments in \cite{A2} to determine the limit of $\Gamma$. Only when the ratio tends to $1$, we know that the limit is $\frac89$. 
In summary, from Figure \ref{fig:Gamma} we see that $\Gamma$ is close to $\frac89$ for large values of $\kappa$ but whether $\Gamma$ can reach $\frac89$ for solutions of the Einstein-Dirac system is an open question. The hope is to answer this analytically in a future investigation. 

We finish this section by also discussing the maximum compactness of boson stars and of Wheeler's idealized spherically symmetric geons.
The question of the compactness of solutions to the Einstein-Klein-Gordon system in the case of $\ell$-boson stars was studied in \cite{AS1}. (For $\ell=0$ the standard boson star first introduced by Kaup \cite{K} is obtained.) In the limit as $\ell\to\infty$, the inequality (\ref{energyconditionOmega}) holds with $\Omega=1$, and $p_r\to 0$, and consequently $\Gamma$ is bounded by $8/9$. The maximum value of the compactness ratio was found to be $0.866$, which is close to $\frac89$. Hence, $\ell$-boson stars behave very similarly regarding their compactness as solutions to the Einstein-Dirac system and to the Einstein\,-Vlasov system. As mentioned in the introduction, there is a similar behaviour for solutions of the Einstein-Maxwell system. Wheeler studied what he calls idealized spherically symmetric geons in \cite{W}. He finds by numerical means that the electromagnetic field is concentrated on a thin shell for the most compact configurations. The numerical value of $\Gamma$ in his case is very close to $8/9$, cf. \cite[p. 525]{W}. It is an interesting question what the common feature of these models is that determines the compactness of the solutions.

\bigskip

\textbf{Acknowledgement: }The authors would like to thank Olivier Sarbach for useful comments.


%
%
%



%

%

\end{document}